\documentclass[notitlepage,aps,onecolumn,accepted=2018-09-18,superscriptaddress]{quantumarticle}

\newif\ifarxiv\arxivtrue
\arxivfalse

\usepackage[margin=2.5cm]{geometry}

\usepackage{amsmath}
\usepackage{amssymb}
\usepackage{graphicx}
\usepackage{hyperref}
\usepackage{color}

\usepackage{xcolor}

\definecolor{ForestGreen}{RGB}{34,139,34}

\hypersetup{
	colorlinks,
	linkcolor={black!30!blue},
	citecolor={red},
	urlcolor={blue}
}

\usepackage[T1]{fontenc}
\usepackage[latin9]{inputenc}

\usepackage{amsthm}
\usepackage{bbm}
\usepackage{mathtools}
\mathtoolsset{centercolon}
\ifarxiv\fi

\usepackage{xspace}

\def\ingraphics#1{\input{#1.tex}}

\usepackage{pgfplots}
\usepgfplotslibrary{patchplots}
\usepackage{tikz}
\usetikzlibrary{arrows}
\usetikzlibrary{decorations.pathreplacing}
\usetikzlibrary{decorations.markings}
\usetikzlibrary{shapes,arrows,positioning,patterns}
\usetikzlibrary{matrix}
\usetikzlibrary{fit}
\usetikzlibrary{fadings}

\def\eat#1{}

\newtheorem{thm}{Theorem}[section]
\newtheorem{defi}[thm]{Definition}
\newtheorem{lem}[thm]{Lemma}
\newtheorem{prop}[thm]{Proposition}
\newtheorem{cor}[thm]{Corollary}

\def\Qref#1{\splitref#1@}
\def\splitref#1:#2@{\thing#1@\;\ref{#1:#2}}
\def\thing#1#2@{\ifx#1sSect.\else\ifx#1fFig.\else\ifx#1TTable\else\ifx#1tThm.\else\ifx#1dDef.\else\ifx#1lLem.\else\ifx#1cCor.\else\ifx#1pProp.\fi\fi\fi\fi\fi\fi\fi\fi}

\def\tr{\operatorname{tr}}
\def\idty{{\leavevmode\rm 1\mkern -5.4mu I}} 

\def\Rl{{\mathbb R}}\def\Cx{{\mathbb C}}
\def\Ir{{\mathbb Z}}\def\Nl{{\mathbb N}}

\def\Grass{Gr}

\def\norm #1{\Vert #1\Vert}

\def\mod{{\mathop{\rm mod}\nolimits}}
\def\ind{{\mathop{\rm ind}\nolimits}\,}

\def\braket#1#2{\langle #1,#2\rangle}
\def\brAket#1#2{\langle #1\vert#2\rangle}

\def\ket #1{\vert#1\rangle}
\def\ketbra #1#2{{\vert#1\rangle\langle#2\vert}}

\def\tr{\mathop{\rm tr}\nolimits}
\def\abs#1{\vert#1\vert}

\let\veps\varepsilon

\def\inv{^{-1}}

\def\re{\Re e}
\def\im{\Im m}

\def\calk#1{\lceil#1\rceil}

\def\dff{\bf} 

\def\BB{{\mathcal B}}{{\def\HH{{\mathcal H}}\def\LL{{\mathcal L}}

\makeatletter
\newcommand{\raisemath}[1]{\mathpalette{\raisem@th{#1}}}
\newcommand{\raisem@th}[3]{\raisebox{#1}{$#2#3$}}
\makeatother

\def\ig{\mathbf I} 
\def\six{\mathop{\mathrm s\mkern-1mu\mathrm i}\nolimits}
\DeclareRobustCommand{\sixnp}{\mathop{\mathrm s\mkern-1mu \textup \i}\nolimits}
\DeclareRobustCommand{\sixR}{\mathop{\hbox{\raisebox{.45em}{$\scriptstyle\rightharpoonup$}}\hspace{-.9em}\sixnp}}
\def\sixL{\mathop{\hbox{\raisebox{.45em}{$\scriptstyle\leftharpoonup$}}\hspace{-.9em}\sixnp}}

\def\sixrel#1:#2{\six(#1{:}#2)}
\def\romkern{\kern-1pt}
\def\rom#1{\textnormal{I\if#11\else\romkern I\if#12\else\romkern I\fi\fi}}

\def\symD{\textnormal{\textsf D}\xspace}
\def\symC{\textnormal{\textsf C}\xspace}

\def\symAIII{\textnormal{\textsf A}\rom3\xspace}
\def\symBDI{\textnormal{\textsf{BD}}\rom1\xspace}

\def\symCII{\textnormal{\textsf C}\rom2\xspace}
\def\symDIII{\textnormal{\textsf D}\rom3\xspace}
\def\symS{\textnormal{\textsf S}\xspace}
\def\igS{\ig(\symS)}   

\def\indF{\ind_{\mathrm F}}

\def\Pl#1{P_{<#1}}
\def\Pg#1{P_{\geq#1}}

\ifarxiv\else\fi

\def\ph{\eta}\def\rv{\tau}\def\ch{\gamma}

\def\Wh{\widehat W} 
\def\Wt{\widetilde W} 

\def\Ah{\widehat A} 
\def\Bh{\widehat B}
\def\Dh{\widehat D}
\def\Fou{{\mathcal F}}
\def\wind{{\mathop{\rm wind}}}

\def\Ba{Q}
\def\pfaff{{\mathop{\rm pf}\nolimits}}

\newcommand{\pf}{\operatorname{pf}}
\newcommand{\sign}{\operatorname{sign}}

\def\antisym{\mathcal A}

\usepackage{hyperref}
\begin{document}

\title{Complete homotopy invariants for translation invariant symmetric quantum walks on a chain
}
\nopagebreak

\author{C. Cedzich}
\affiliation{Institut f\"ur Theoretische Physik, Leibniz Universit\"at Hannover, Appelstr. 2, 30167 Hannover, Germany}
\affiliation{Institut f\"ur Theoretische Physik, Universit\"at zu K\"oln, Z\"ulpicher Stra{\ss}e 77, 50937 K\"oln, Germany}
\author{T. Geib}
\affiliation{Institut f\"ur Theoretische Physik, Leibniz Universit\"at Hannover, Appelstr. 2, 30167 Hannover, Germany}
\author{C. Stahl}
\affiliation{Institut f\"ur Theoretische Physik, Leibniz Universit\"at Hannover, Appelstr. 2, 30167 Hannover, Germany}
\author{L. Vel\'azquez}
\affiliation{Departamento de Matem\'{a}tica Aplicada \& IUMA,  Universidad de Zaragoza,  Mar\'{\i}a de Luna 3, 50018 Zaragoza, Spain}
\author{A.~H. Werner}
\affiliation{{QMATH}, Department of Mathematical Sciences, University of Copenhagen, Universitetsparken 5, 2100 Copenhagen, Denmark,}
\affiliation{{NBIA}, Niels Bohr Institute, University of Copenhagen, Denmark}
\author{R.~F. Werner}
\affiliation{Institut f\"ur Theoretische Physik, Leibniz Universit\"at Hannover, Appelstr. 2, 30167 Hannover, Germany}

\begin{abstract}
We provide a classification of translation invariant one-dimensional quantum walks with respect to continuous deformations preserving unitarity, locality, translation invariance, a gap condition, and some symmetry of the tenfold way. The classification largely matches the one recently obtained (arXiv:1611.04439) for a similar setting leaving out translation invariance. However, the translation invariant case has some finer distinctions, because some walks may be connected only by breaking translation invariance along the way, retaining only invariance by an even number of sites. Similarly, if walks are considered equivalent when they differ only by adding a trivial walk, i.e., one that allows no jumps between cells, then the  classification collapses also to the general one. The indices of the general classification can be computed in practice only for walks closely related to some translation invariant ones. We prove a completed collection of simple formulas in terms of winding numbers of band structures covering all symmetry types. Furthermore, we determine the strength of the locality conditions, and show that the continuity of the band structure, which is a minimal requirement for topological classifications in terms of winding numbers to make sense, implies the compactness of the commutator of the walk with a half-space projection, a condition which was also the basis of the general theory. In order to apply the theory to the joining of large but finite bulk pieces, one needs to determine the asymptotic behaviour of a stationary Schr\"odinger equation. We show exponential behaviour, and give a practical method for computing the decay constants.
\end{abstract}


\maketitle


\section{Introduction}
The topological classification of quantum matter has recently attracted a lot of attention \cite{kitaevPeriodic,KitaevLectureNotes,MPSphaseII,Schnyder1,Schnyder2,HasanKaneReview,KaneMeleQSH,KaneMeleTopOrder,ZhangTopologicalReview}. It seems that this perspective provides us with some system properties which are stable beyond expectations. Whatever more detailed question one might ultimately wish to investigate, understanding this classification is as essential as a first step as knowing about continents before studying a map in an atlas. The basic pattern of classification, and at the same time the guarantee for its stability, is typically homotopy theory. That is, we consider two systems to be equivalent, or ``in the same topological phase'', if one can be deformed continuously into the other while retaining some key properties. The classification depends critically on these constraints. The typical ones are: discrete symmetries, a spectral gap and some finite range condition (locality). In the discrete time setting, where dynamics is in discrete steps, unitarity is a further constraint. For the description of bulk matter, translation invariance is also often imposed. For translation invariant systems one can go to momentum space, where the unitary time step or, in the continuous time case the Hamiltonian can be diagonalized for each quasi-momentum separately, leading to a vector bundle over the Brillouin zone. Locality amounts to a smoothness condition for this bundle, and symmetry and gap conditions also translate easily. However, in this setting some of the most interesting phenomena are absent, namely all those relating to boundary effects, when two bulk systems in different phases are joined. Folklore has it that when these bulk phases can only be joined by closing a spectral gap on the way, then the gap must ``close at the boundary'', i.e., we expect eigenvalues or modes at the boundary. Plausible as this may seem, proving such statements is another matter. Since it involves a non-translation invariant situation, the vector bundle classification is no longer available, and while translation invariant systems (say, with finite range) are described by finitely many parameters, the space of possible deformations becomes vastly larger.

For the case of one dimensional discrete time systems (``quantum walks'') \cite{Grimmet,electric,dynlocalain,QDapproach,TRcoin} we have provided a classification in terms of three independent indices $(\sixL(W),\six_-(W),\sixR(W))$ taking values in a certain index group depending on the symmetries imposed \cite{LongVersion}. The classification makes no assumption on translation invariance and is shown to be complete in the sense that two walks can be deformed into each other if and only if these indices coincide. This classification is mostly in agreement with the intuitions of the heuristic literature about translation invariant systems \cite{Kita,Kita2,Asbo1,Asbo2,Asbo3}, where these triples are of the form $(-n,0,n)$. It also sharpens the stability properties of the classification. In particular, in contrast to the continuous time case, there are walks which differ only in a finite region, but can still not be deformed into each other, i.e., local perturbations are not necessarily ``gentle''. The index $\six_-(W)$ measures exactly this difference.

To our surprise we found that some aspects of the seemingly simpler, translation invariant subcase presented additional difficulties. First, one would like to have a straightforward formula for the indices in the translation invariant case. These formulas existed in the solid state literature, but we had to connect them to the general classification of walks in \cite{LongVersion}.

Second, the completeness problem had to be addressed (\Qref{sec:classification}). The  theory in \cite{LongVersion} asserts a complete classification, meaning that walks with equal indices can actually be deformed into each other. But when these walks are translation invariant, it is not a priori clear whether this deformation can be chosen to preserve translation invariance as well. In fact, we show below that this is only true for two of the five symmetry types with non-trivial invariant. For the remaining three we introduce an additional binary homotopy invariant, which together with the index from \cite{LongVersion} is complete. The additional invariant turns out to be unstable in several ways: It trivializes, when we allow regrouping of neighbouring cells, i.e., breaking the translation invariance to invariance only under even translations. The same happens if we allow addition of trivial walks (those that act separately and equally in every cell), as is customary in ``stable'' homotopy. Thus allowing either stabilizing with trivial additions or period-2 regrouping the index from \cite{LongVersion} is complete even in the translation invariant setting.

Third, the general theory used a very weak locality condition, allowing the amplitudes for long jumps to decay like $\abs x^{-1-\veps}$. In the translation invariant setting the decay of jump amplitudes should translate into a smoothness property of the band structure, but which? We show in \Qref{sec:essloc} that, while locality in the sense of a norm limit of banded matrices translates exactly into continuity, the locality condition of \cite{LongVersion} translates to the weaker property ``quasi-continuity'' of the bands. It is a challenge to state index formula in this setting. We carefully state these as winding numbers, which only require continuity, not differentiability of the walk unitary as a function on the Brillouin zone (the most common case). In the quasi-continuous case this would be replaced by radial limits of winding numbers of analytic functions with a discontinuous boundary behaviour.

Finally, we had discussed in \cite{LongVersion} how the theory, which is set up for infinite systems, actually applies ``approximately'' to finite systems. Would protected boundary eigenvalues appear near the interface between two long but finite intervals of distinct bulk phases? The crucial quantity here is the exponential behaviour of solutions of the stationary Schr\"odinger equation for the translation invariant versions of the bulk. This is a system of linear recursion relations, which cannot in general be solved for the highest term, so the standard theory of exponential behaviour of solutions to finite difference equations does not immediately apply. Nevertheless, exponential behaviour is shown in full generality in \Qref{sec:decay}.
 
In several contexts (e.g., \cite{LongVersion,OldIndex}) we have seen that walks exhibit non-trivial additional invariants compared to the Hamiltonian case. However, in the translation invariant case the difference is not so pronounced. The essential gap condition then implies that the ``effective Hamiltonian'' $H_{\rm eff}=i \log W$ is well-defined with a branch cut on the negative real axis. This provides a continuous (hence homotopy compatible) mapping between the discrete and the continuous time case. Therefore, Theorems from the literature on Hamiltonian systems can be taken over. Conversely, our results of Sects.~\ref{sec:essloc} and \ref{sec:classification} apply to the Hamiltonian case without further ado.

\subsection{Connection to the literature}

Topological phenomena are well-studied in many areas of physics, and consequently there are many settings in which related questions are investigated, some belonging to different disciplines and scientific cultures. Therefore, we give a perhaps unusually extended (and slightly subjective) description of known and potential connections, including pointers to the literature. We are fully aware that we cannot give full justice to this rich field, and apologize for any serious omissions.

Our own point of departure was in the quantum information community (mathematical physics division). For many years the Hannover group has enjoyed a close collaboration with the experimental group in Bonn \cite{decoherence,molecules,electric}, who realize quantum walks of neutral atoms in an optical lattice. Naturally, this has partly determined our framework. Interest in this community was initially in long time asymptotics (ultimately with algorithms in mind) and in recent years in the possibility to recast some phenomena of interest in solid state physics for quantum simulation in a completely controlled environment.

Some of the main connections of our work to work done in related fields are the following:

\begin{itemize}
\item Most of the physics literature on topological phases is for free fermion systems in the continuous time (Hamiltonian) case \cite{Altland-Zirnbauer,heinzner2005symmetry,Zirnbauer-Symmetry,KitaevMajorana,kitaevPeriodic,KitaevLectureNotes,Schnyder1,Schnyder2,roy2008topological,qi2010topological,PhysRevLett.102.187001,hastings2013classifying}. In our paper the Hilbert space is a straightforward abstracted version of a system  Hilbert space with basis vectors labelled by the discrete position and internal excitations of a walking atom. In contrast, for a free fermion system it is a space of modes of elementary excitations of a many particle system, whose basis vectors label creation/annihilation of both particles and holes (Nambu space). Therefore the particle-hole symmetry is built into the structure from the outset, whereas it may or may not hold in our setting.

\item The solid state literature is rather uneven in its level of rigour. Also, as we found in retrospect, argumentation for general statements is not emphasized, and often not given in detail, although the authors could have done so. We are far from criticizing this, but, as a complementary approach follow a different tack, in order to achieve the kind of clarity we require as mathematical physicists, and in the process to spare mathematically inclined readers some frustration. We are also in a collaboration \cite{AsbothConnect} with a colleague closer to the solid state community, to straighten out conceptual differences.

\item We are working in the discrete time case, with a single unitary operator $W$ defining the dynamical step. This often arises in the Floquet setting, in which $W$ is one period of a periodically driven Hamiltonian system. Of course, it can be argued that the experimental system of our partners is precisely of this type. In previous work \cite{OldIndex,LongVersion} we found additional topological invariants in comparison to the Hamiltonian case, i.e.\ a non-trivial interplay of the locality, gap and unitarity requirements. The difference in the current paper, which is concerned with the translation invariant case only, is not so great. In fact, the basic index formulas  (\eqref{preberry}, \eqref{Berryintegral}, \eqref{windingcontinuous}, \eqref{windingintegral}, \eqref{DIIIberry})  can be found in the literature (we thank a referee for tracking \eqref{DIIIberry} in \cite{qi2010topological}), as applied to the ``effective Hamiltonian'' $H_{\rm eff}=-i\log{W}$. This works in the translation invariant case, because there can be no eigenvalues in the spectral gap of $W$ at $-1$, so there are no problems with the branch cut of the logarithm. Yet, we choose to work on the level of unitaries, also to establish a direct connection to the general theory in \cite{LongVersion}.

\item These formulas are not the main result of our paper. In fact, we have been aware of (most of) them since the beginning of our project, when we took inspiration from the first papers concerned with the topological classification of walks \cite{Kita,Kita2,KitaObservation,Asbo1,Asbo2,Asbo3,obuse2015unveiling}. Our point here is rather more subtle, namely to explore the ``completeness'' of these invariants: Thus equality of invariants of two walks should imply that they can be deformed into each other, while keeping the relevant structures (gap, symmetry, locality) intact. Proving completeness is the way to show that the classification job has been finished, and no hidden obstructions have been overlooked. However, the completeness question is very sensitive to a change of setting. On the one hand, it is surprising that the complete classification of non-translation invariant walks \cite{LongVersion} is essentially determined by the (prima facie) much simpler translation invariant version of the question: Every class can be realized by a local perturbation of  two translation invariant walks cut in half and joined together. On the other hand, new classes might arise in the translation invariant case, because two such walks could be deformable into each other, {\it but only} by breaking the translational symmetry. In fact, as we show in this paper, this does happen for the symmetry classes \symD\ (\Qref{prop:symDti}), \symBDI\ (\Qref{lem:BDIcomplete}), but not for \symDIII\ (\Qref{prop:D3complete}), \symCII (\Qref{lem:CIIcomplete}), and \symAIII (see beginning of \Qref{sec:chiralcomplete}). However, all additional invariants disappear if we allow grouping of two neighbouring cells, i.e., deformations which break translational symmetry and are only invariant under even translations.

\item There is an interesting alternative way to look at  Floquet systems with symmetry. What we are considering is only the unitary step operator $W$, i.e., the result of Hamiltonian driving over a whole period, and the classification of such operators. However, as pointed out to us by J.~Asboth, one could also be interested in the homotopy classification of the entire driving process, or ``protocol''. Of course, in a setting with discrete symmetries this only really makes sense, when the driving process satisfies the symmetry along the whole way. With time symmetric driving this gives  a special role to the unitary operator at half-period. In the experimental system we had in mind this symmetry in the driving process is definitely not respected. Rather, ``topological protection'' is engineered in the same sense as ``overprotection of children'': many unintended influence can ruin it. This is (partly) different in the free fermion case: One important symmetry (particle-hole) is satisfied anyway, and also time reversal is sometimes naturally obeyed in the absence of magnetic fields. A clear recent indication that discussing the protocol rather than the overall Floquet operator $W$ came in \cite{harperroy2017periodictable,harperroy2018chiral}. They study ``loops'' for which $W=\idty$, and get non-trivial classification. Of interest is also  \cite{Rudner}, and we are in the process of looking for further possibilities \cite{AsbothConnect}.

\item In contrast to much of the literature, our paper concerns only 1D lattice systems. This may look very constrained, but on the other hand, we allow not just finite jumps, but give a formulation that allows the longest possible tails for the jump amplitudes. This is closely related to notions of locality \cite{graf2018bulk} that will be crucial for finding an efficient theory in higher dimension. Locality as a kind of bound on the jumps allowed in a single walk step is well captured by a so-called {\it coarse structure} \cite{RoeCoarse}, and its associated Roe C*-algebra. The main natural choices in one-dimension relate to either uniformly finite jump lengths (banded matrices) or to finite, but possibly site dependent jump lengths, which amounts to a compactness condition (``essential locality'') introduced in \cite{LongVersion}. It would seem plausible that in the translation invariant case these coincide, but as we show in \Qref{sec:essloc} a subtlety remains, so that we find (\Qref{prop:quasiConti})that certain translation invariant walks with discontinuous band structure are nevertheless susceptible to the classification of \cite{LongVersion}. We also take some care to formulate index formulas for merely continuous band structures, i.e., in a purely topological way, not requiring derivatives.

\item
In the mathematical literature the classification of topological insulators is usually done in terms of K-theory \cite{kitaevPeriodic,Prodan,SchulzFloquet,SchulzZ2,SchulzBuch,harperroy2017periodictable,KubotaCoarse,MeyerCoarse}. A built-in feature of K-theory is homotopy stabilization, by which addition of trivial systems is considered as a neutral operation not changing the K-group classification. That this stabilization makes a difference for walks is demonstrated by the additional invariants we derive in our setting. These are ``unstabilized homotopy'' invariants, but disappear, when added trivial systems are allowed to be included in a homotopy path \Qref{sec:stabelHomo}.

\end{itemize}

\subsection{Recap of the general theory}\label{sec:nonti}

Before delving into the translation invariant case, let us provide a brief overview of the setting and results of our index theory for symmetric quantum walks in one spatial dimension \cite{LongVersion}. We introduce the systems we consider, attribute topological invariants and discuss their stability under different classes of perturbations. Also, we review some of the hallmark results of \cite{LongVersion} like the gentle decoupling theorem and the completeness of the invariants.

The Hilbert spaces of the quantum systems described by the theory in \cite{LongVersion} are assumed to have the form $\HH=\bigoplus_{x\in\Ir}\HH_x$, where each $\HH_x$ is a finite-dimensional Hilbert space which we refer to as the \textbf{cell} at position $x$. For $a\in\Ir$ we denote by $\Pg a$ the projection onto the subspace $\bigoplus_{x\geq a}\HH_x$ and often abbreviate $P=\Pg0$.

On $\HH$, we consider discrete, involutive \textbf{symmetries} which act locally in each cell, i.e., unitary or anti-unitary operators which, by Wigner's theorem, square to a phase times the identity and which commute with each $\Pg a$. The set of such operators forms an abstract group which in the setting we consider consists either only of the identity, the identity and a single involution or the Klein four-group. In particular, we always choose trivial multiplication phases for the symmetries, i.e. we assume that they commute exactly. Apart from their square and their unitary or antiunitary character, the symmetries are distinguished by their relations with ``symmetric'' operators, i.e. unitary or hermitean operators $W$ or $H$ that ``satisfy the symmetries''. Concretely, we restrict our attention to the following symmetries which are frequently considered in solid-state physics:
\begin{itemize}
	\item[] {\dff particle-hole symmetry} $\ph$, which is antiunitary satisfying
	$\ph W\ph^*=W$, resp.\ $\ph H\ph^*=-H$,
	\item[] {\dff time reversal symmetry} $\rv$, which is antiunitary satisfying
	$\rv W\rv^*=W^*$, resp.\ $\rv H\rv^*=H$,
	\item[] {\dff chiral symmetry} $\ch$, which is unitary satisfying
	$\ch W\ch^*=W^*$ resp.\ $\ch H\ch^*=-H$.
\end{itemize}
For each group of symmetries, the information on the square and the unitary or antiunitary character of each symmetry together with the relations with symmetric operators constitute a \textbf{symmetry type}. These symmetry types establish the so-called \textit{tenfold way} \cite{Altland-Zirnbauer}.
A unitary or hermitean operator on $\HH$ is called \textbf{admissible} for a symmetry type, if it satisfies the relations of the symmetry type and is \textbf{essentially gapped}, i.e., possesses only finitely degenerate eigenvalues in a small neighborhood around the symmetry-invariant points $\pm1$ in the unitary and $0$ in the hermitean case. A key assumption for the classification is, that the symmetries are \textbf{balanced} locally, i.\ e.\ in each cell there exists a strictly gapped, admissible unitary (Hermitian) \cite{LongVersion}.

The physical systems we are concerned with in \cite{LongVersion} and in the current paper are \textbf{quantum walks}, i.e. unitary operators $W$ that additionally satisfy a locality condition. In many earlier papers, this locality condition is formulated as a finite upper bound on the jump length, in which case we call $W$ \textbf{strictly local}. If, more generally, $\Pg a-W^*\Pg aW$ is a compact operator for all $a$, we call $W$ \textbf{essentially local}, and this standing assumption from \cite{LongVersion} will also be true for all walks in this paper (see, however, \Qref{sec:essloc}).

Using elementary group theory we assign an abelian symmetry group $\igS$ to each symmetry type $\symS$ \cite{LongVersion}. The symmetry types with $\igS$ non-trivial are displayed in Table \ref{Tab:sym}.
Vaguely speaking, a topological classification then assigns to each system admissible for a symmetry type $\symS$ an element of the corresponding index group $\igS$, such that this classification
is stable under a certain class of perturbations.
To find such a classification for admissible unitaries, we first realize that the eigenspaces at $\pm1$ of such operators are invariant under the action of the symmetries and therefore play a special role:
we assign elements $\six_+(W)$ and $\six_-(W)$ of $\igS$ to the $+1$- and $-1$-eigenspaces of unitaries $W$ admissible for $\symS$, respectively. An analogous construction for admissible Hermitean operators assigns an element $\six(H)\in\igS$ to the symmetry-invariant $0$-eigenspace of $H$. Basic properties of these \textbf{symmetry indices} are that they are additive with respect to direct sums of operators and prove to be invariant under \textbf{gentle perturbations}, i.e., they are constant along norm-continuous paths $t\mapsto W_t$ for which every $W_t$ is admissible.

The sum $\six(W)=\six_+(W)+\six_-(W)$ is called the symmetry index of $W$, and one quickly argues that, in contrast to $\six_\pm$, $\six(W)$ is invariant not only under gentle but also under \textbf{compact perturbations}, i.e. admissible operators $W'$ for which $W-W'$ is compact. This theory of classifying admissible unitary operators by the symmetry indices $\six_\pm$ works independently of any cell structure of the Hilbert space $\HH$.

Taking into account such a cell structure, one can argue that physically relevant perturbations are local, i.e. supported on a finite number of cells. It is therefore important to find a classification of physical systems which is stable under such local perturbations. In the Hamiltonian case this is straightforward, since every compact and therefore every local perturbation $H'$ of $H$ is automatically gentle along the path $t\mapsto tH'+(1-t)H$. The unitary case, however, is more involved as one can easily construct compact perturbations of a given unitary operator which are not gentle.

To nevertheless find such a classification for admissible walks, in \cite{LongVersion} we introduced further invariants which are defined in terms of the cell structure of $\HH$: even though not unitary, by the locality condition the operators $\Pg a W \Pg a$ and $\Pl a W \Pl a$ are \textbf{essentially unitary} on the respective half chains, i.e. $W^*W-\idty$ and $WW^*-\idty$ are compact on $\Pg a\HH$ and $\Pl a\HH$. Realizing that the imaginary part of an essentially unitary operator is hermitean, we define the invariants $\sixR(W):=\six(\Pg a W \Pg a)\equiv\six(\im(\Pg a W \Pg a))$ with $a\in\Ir$, and similarly for $\sixL(W)$.
They are independent of $a$ and invariant with respect to gentle and compact perturbations. Furthermore, like for the ``spectral'' invariants $\six_\pm$ we have $\six(W)=\sixL(W)+\sixR(W)$. An overview of the indices defined in \cite{LongVersion} is provided in Table \ref{tab:indices}.

\begin{table}
  \begin{center}
    \begin{tabular}{cc|c}
     $\six_+(W_L)$&$\six_+(W_R)$&$\six_+(W)$\\
     $\six_-(W_L)$&$\six_-(W_R)$&$\six_-(W)$\\\hline
     $\sixL (W)$&$\sixR (W)$&$\six (W)$\vrule height 5pt depth 0pt width 0pt
    \end{tabular}
  \end{center}
  \caption{\label{tab:indices}A summary of the symmetry indices of a gently decoupled walk $W=W_L\oplus W_R$ as introduced in \cite{LongVersion}. Here, the quantities in the upper left quadrant are neither invariant under gentle nor under compact perturbations whereas their row and column sums in the bottom row and the right column are.}
\end{table}

These ``spatial'' symmetry indices allow to prove a \textbf{bulk-boundary-correspondence}: whenever two translation invariant bulks $W_L$ and $W_R$ are in different topological phases, a crossover walk $W$ between them hosts topologically protected eigenfunctions.
The number of these eigenfunctions is lower bounded by $\abs{\six(W)}=\abs{\sixR(W_L)-\sixR(W_R)}$ and, as we prove in \Qref{sec:decay} of the present paper, for translation invariant walks they are exponentially localized near the boundary.

Another way to split an admissible walk $W$ into a left and a right half $W_L$ and $W_R$, respectively, is to decouple gently, i.e. to find a gentle perturbation of $W$ such that $W'=W_L\oplus W_R$. The existence of such decouplings is guaranteed by the \textbf{gentle decoupling theorem} for which a necessary and sufficient condition is the vanishing of the Fredholm index of $PWP$, a condition satisfied by every admissible unitary
\cite{LongVersion}. Moreover, due to its gentleness, such a decoupling preserves all symmetry indices discussed so far, whereas $W\mapsto PWP+(\idty-P)W(\idty-P)$ preserves only $\sixL(W),\sixR(W)$ and $\six(W)$.

This existence of gentle decoupling plays an essential role when discussing the \textbf{completeness} of the topological invariants: having proved that all symmetry indices are invariant under gentle perturbations evokes the question, whether the converse is also true, i.e. whether two systems which share the same values for the symmetry indices can always be deformed into each other along an admissible path. Clearly, the set of symmetry indices suitable for a classification depends on the perturbations one allows for. In \cite{LongVersion}, we answered the completeness question in the affirmative for the following settings:
\begin{itemize}
	\item[(\rom1)] All walks with respect to gentle perturbations and independent invariants $\sixR,\sixL,\six_-$
	\item[(\rom2)] All walks with respect to both gentle and compact perturbations and independent invariants $\sixR,\sixL$.
	\item[(\rom3)] All admissible unitaries with respect to gentle perturbations and independent invariants $\six_+,\six_-$.
\end{itemize}

In this paper we restrict our attention to translation invariant systems. For such systems, the gap condition implies that the indices $\six_\pm(W)$ and therefore also $\six(W)$ vanish. Thus, $\sixR(W)$ or, equivalently, $\sixL(W)$ provides the full information on the classification of such systems.

\begin{table}
	\begin{center}
		\begin{tabular}{|c||c|c|c||c||c|c|}
			$\symS$  &$\ph^2$  &$\rv^2$  &$\ch^2$  &$\igS$ &$\six$	\\
			\hline
			\symD    &$\idty$  &         &         &$\Ir_2$  &$d\,\mod2$  \\[3pt]
			\symAIII &         &         &$\idty$  &$\Ir$    &$\tr\ch$
			\\[3pt]
			\symBDI  &$\idty$  &$\idty$  &$\idty$  &$\Ir$    &$\tr\ch$
			\\[3pt]
			\symCII  &$-\idty$ &$-\idty$ &$\idty$  &$2\Ir$   &$\tr\ch$
			\\[3pt]
			\symDIII &$\idty$  &$-\idty$ &$-\idty$ &$2\Ir_2$ &$d\,\mod4$
		\end{tabular}\hspace{10pt}
		\caption{\label{Tab:sym}The non-trivial symmetry types of the tenfold way where the first column gives the label of the Cartan classification \cite{Altland-Zirnbauer}. The index groups $\igS$ are calculated explicitly in \cite{LongVersion}. The last column gives an explicit expression for the symmetry index of finite-dimensional realizations of the symmetry type.}
	\end{center}
\end{table}

\subsection{Translation invariance and band structure}\label{sec:bands}

In the translation invariant case the cells $\HH_x$ all need to be identical so that $\HH=\bigoplus\HH_x\cong \ell^2(\Ir)\otimes\HH_0$. We will assume that the symmetry operators in each cell are likewise all equal. Then, an operator $W\in\BB(\HH)$ is called \textbf{translation invariant} if it commutes with the translations or, equivalently, if
\begin{equation}\label{Wti}
  (W\psi)(x)=\sum_yW(y)\psi(x-y)
\end{equation}
for some operators $W(y):\HH_0\to\HH_0$. For strictly local walks only finitely many $W(y)$ are non-zero. The above convolution can be turned into a product by the Fourier transform
$\Fou:\ell^2(\Ir)\to\LL^2([-\pi,\pi],dk/(2\pi))$ with the convention
\begin{equation}\label{Fou}
  (\Fou\psi)(k)=\sum_xe^{ikx}\psi(x).
\end{equation}
We use the same notation for $\Fou\otimes\idty_d$, or $\HH_0$-valued $\psi$. Of course, the interval $[-\pi,\pi]$, or Brillouin zone, or momentum space, just serves as a parametrization of the circle, and the Fourier transforms of sufficiently rapidly decaying $\psi$ will be continuous with periodic boundary conditions.
Then, the walk is determined by the function $k\mapsto \Wh(k)\in U(d)$ via
\begin{equation}\label{FouW}
  (\Fou W\psi)(k)=\Wh(k)(\Fou\psi)(k)\ ,\quad \Wh(k)=\sum_xe^{ikx}W(x)
\end{equation}
For a strictly local walk $\Wh$ is a matrix valued Laurent polynomial in the variable $\exp(ik)$. For rapidly decreasing $W(x)$ we get smooth $\Wh$, see \Qref{sec:essloc} for more details.

A complete homotopy classification that does not demand any symmetries or gap conditions is provided by the index ($\ind$) of a quantum walk, thoroughly studied in the case of strictly local walks in \cite{OldIndex} and extended to essentially local walks in \cite{LongVersion}. In this general case the alluded index is given in terms of the Fredholm index ($\ind_F$) of $PWP$, in fact $\ind(W)=-\ind_F(PWP)$.
For translation invariant $W$ this becomes the total winding number of the quasi-energy spectrum around the Brillouin zone:
\begin{defi}\label{def:oldindex}
	Let $W$ be a translation invariant quantum walk with Fourier transform $\Wh(k)$. Then, its {\bf index} is given by the winding number of $\det\Wh(k)$ around the origin, which in the case of a continuously differentiable $\Wh(k)$ is given by
	\begin{equation}\label{eq:indW}
	\ind(W)=\frac{1}{2\pi i}\int_{-\pi}^\pi\mskip-10 mu dk\left(\frac{\partial}{\partial k} \log\det\Wh(k)\right).
	\end{equation}
\end{defi}
Yet, for the theory we develop in this paper, $\ind(W)$ automatically vanishes by the existence of essential gaps in the spectrum \cite[Proposition VII.1]{LongVersion}.

Now, turning to the classification of translation invariant walks with symmetries, we need to express the symmetry conditions in terms of $\Wh$. These are then given by
\begin{equation}\label{Fousym}
\ph\Wh(k)\ph^*=\Wh(-k)    ,\ \qquad
\rv\Wh(k)\rv^*=\Wh(-k)^*  ,\ \mbox{and} \qquad
\ch\Wh(k)\ch^*=\Wh(k)^*,
\end{equation}
where the operators $\ph,\rv,\ch$ denote the one-cell symmetries acting on $\HH_0\simeq\Cx^d$. Note that the anti-unitarily implemented symmetries flip the sign of $k$. This is necessary to guarantee that the position basis elements, i.e. $\exp(i k)$, are invariant under the symmetries.

As a finite-dimensional, $k$-dependent unitary matrix, $\Wh(k)$ can be diagonalized, i.e.
\begin{equation}\label{Whk}
	\Wh(k)=\sum_{\alpha=1}^d e^{i\omega_\alpha(k)}\Ba_\alpha(k),
\end{equation}
where $\omega_\alpha(k)$ are the quasi-energies and $\Ba_\alpha$ the band projections.

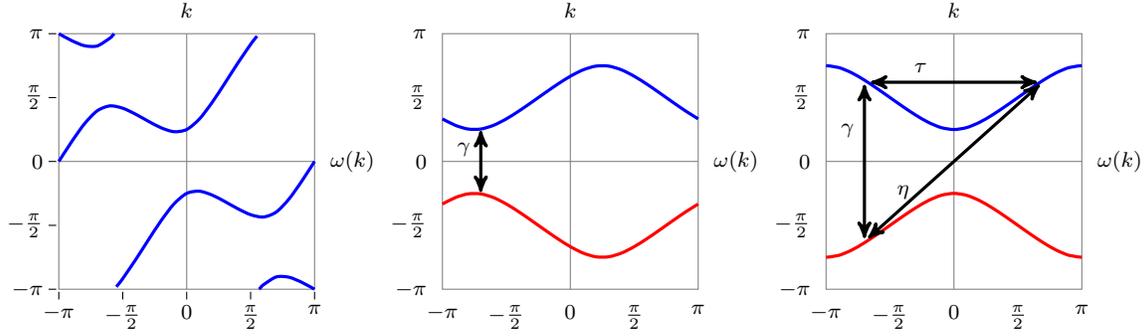
\begin{figure}
    \begin{center}
		\tikzset{
	>=stealth',
	mid graphic/.style={
		xshift=3cm
	},
	right graphic/.style={
		xshift=6cm
	}
}

\begin{tikzpicture}
	[
scale=1.7,
font=\footnotesize
]

	\draw[gray] (-1.,-1.)  rectangle +(2,2);
	\draw[gray] (-1,0) -- +(2,0);
	\draw[gray] (0,-1) -- +(0,2);
	\foreach \i in {-1,-.5,0,.5,1}{
		\draw[align=left] (-1.02,{\i}) -- (-1.08,{\i});
		\draw[align=left] ({\i},-1.02) -- ({\i},-1.08);
	}
	
	\draw (-1.05,-1)  node[left,align=left]{ $-\pi$};
	\draw (-1.05,-.5)  node[left,align=left]{ $-\frac{\pi}{2}$};
	\draw (-1.05,0)  node[left,align=left]{ $0$};
	\draw (-1.05,.5)  node[left,align=left]{ $\frac{\pi}{2}$};
	\draw (-1.05,1)  node[left,align=left]{ $\pi$};
	
	\draw (-1,-1.05)  node[below,align=center]{ $-\pi$};
	\draw (-.5,-1.05)  node[below,align=center]{ $-\frac{\pi}{2}$};
	\draw (0,-1.05)  node[below,align=center]{ $0$};
	\draw (.5,-1.05)  node[below,align=center]{ $\frac{\pi}{2}$};
	\draw (1,-1.05)  node[below,align=center]{ $\pi$};

	\draw (0,1.05) node[above,align=center]{ $k$};
	\draw (1.05,0) node[right,align=left]{ $\omega(k)$};

	\draw[very thick, blue,smooth] plot coordinates {(-1., 0) (-0.9, 0.154026) (-0.8, 0.293856) (-0.7, 0.396105) (-0.6, 0.434779) (-0.5, 0.416667) (-0.4, 0.370119) (-0.3, 0.314718) (-0.2, 0.263562) (-0.1, 0.23304) (0., 0.25) (0.1, 0.33304) (0.2, 0.463562) (0.3, 0.614718) (0.4, 0.770119) (0.5,0.916667) (0.516667, 0.939036) (0.533333, 0.960523) (0.55, 0.980991)};
	
	\draw[ very thick, blue,smooth] plot coordinates {(0.566667, -0.999706) (0.583333, -0.981725) (0.6, -0.965221) (0.7, -0.903895) (0.8, -0.906144) (0.9, -0.945974) (1., -1)};

	\draw[very thick, blue,smooth] plot coordinates { (-0.55, -0.980991) (-0.533333, -0.960523) (-0.516667, -0.939036) (-0.5, -0.916667) (-0.4, -0.770119) (-0.3, -0.614718) (-0.2, -0.463562) (-0.1, -0.33304) (0., -0.25) (0.1, -0.23304) (0.2, -0.263562) (0.3, -0.314718) (0.4, -0.370119) (0.5, -0.416667) (0.6, -0.434779) (0.7, -0.396105) (0.8, -0.293856) (0.9, -0.154026) (1., 0)};
	
	\draw[very thick, blue,smooth] plot coordinates {(-1., 1) (-0.9, 0.945974) (-0.8, 0.906144) (-0.7, 0.903895) (-0.6, .965221) (-0.583333, 0.981725) (-0.566667, 0.999706)};


	\draw[mid graphic,gray] (-1.,-1.)  rectangle +(2,2);
	\draw[mid graphic,gray] (-1,0) -- +(2,0);
	\draw[mid graphic,gray] (0,-1) -- +(0,2);
	
	\draw[mid graphic] (-1.05,-1)  node[left,align=left]{ $-\pi$};
	\draw[mid graphic] (-1.05,-.5)  node[left,align=left]{ $-\frac{\pi}{2}$};
	\draw[mid graphic] (-1.05,0)  node[left,align=left]{ $0$};
	\draw[mid graphic] (-1.05,.5)  node[left,align=left]{ $\frac{\pi}{2}$};
	\draw[mid graphic] (-1.05,1)  node[left,align=left]{ $\pi$};

	\draw[mid graphic] (-1,-1.05)  node[below,align=center]{ $-\pi$};
	\draw[mid graphic] (-.5,-1.05)  node[below,align=center]{ $-\frac{\pi}{2}$};
	\draw[mid graphic] (0,-1.05)  node[below,align=center]{ $0$};
	\draw[mid graphic] (.5,-1.05)  node[below,align=center]{ $\frac{\pi}{2}$};
	\draw[mid graphic] (1,-1.05)  node[below,align=center]{ $\pi$};

	\draw[mid graphic] (0,1.05) node[above,align=center]{ $k$};
	\draw[mid graphic] (1.05,0) node[right,align=left]{ $\omega(k)$};

	\draw[mid graphic, very thick, blue,smooth] plot coordinates {(-1., 0.333333) (-0.9, 0.28304) (-0.8, 0.253895) (-0.7, 0.253895) (-0.6, 0.28304) (-0.5, 0.333333) (-0.4, 0.395974) (-0.3, 0.464718) (-0.2, 0.535282) (-0.1, 0.604026) (0., 0.666667) (0.1, 0.71696) (0.2, 0.746105) (0.3, 0.746105) (0.4, 0.71696) (0.5, 0.666667) (0.6, 0.604026) (0.7, 0.535282) (0.8, 0.464718) (0.9, 0.395974) (1., 0.333333)};
	
	\draw[mid graphic, very thick, red,smooth] plot coordinates {(-1., -0.333333) (-0.9, -0.28304) (-0.8, -0.253895) (-0.7, -0.253895) (-0.6, -0.28304) (-0.5, -0.333333) (-0.4, -0.395974) (-0.3, -0.464718) (-0.2, -0.535282) (-0.1, -0.604026) (0., -0.666667) (0.1, -0.71696) (0.2, -0.746105) (0.3, -0.746105) (0.4, -0.71696) (0.5, -0.666667) (0.6, -0.604026) (0.7, -0.535282) (0.8, -0.464718) (0.9, -0.395974) (1., -0.333333)};
	
	\draw[mid graphic,very thick, black,<->] (-0.7, 0.24) -- (-0.7, -0.24) node[pos=0.3,left]{$\ch$};

	\draw[right graphic,gray] (-1.,-1.)  rectangle +(2,2);
	\draw[right graphic,gray] (-1,0) -- +(2,0);
	\draw[right graphic,gray] (0,-1) -- +(0,2);

	\draw[right graphic] (-1.05,-1)  node[left,align=left]{ $-\pi$};
	\draw[right graphic] (-1.05,-.5)  node[left,align=left]{ $-\frac{\pi}{2}$};
	\draw[right graphic] (-1.05,0)  node[left,align=left]{ $0$};
	\draw[right graphic] (-1.05,.5)  node[left,align=left]{ $\frac{\pi}{2}$};
	\draw[right graphic] (-1.05,1)  node[left,align=left]{ $\pi$};
	
	\draw[right graphic] (-1,-1.05)  node[below,align=center]{ $-\pi$};
	\draw[right graphic] (-.5,-1.05)  node[below,align=center]{ $-\frac{\pi}{2}$};
	\draw[right graphic] (0,-1.05)  node[below,align=center]{ $0$};
	\draw[right graphic] (.5,-1.05)  node[below,align=center]{ $\frac{\pi}{2}$};
	\draw[right graphic] (1,-1.05)  node[below,align=center]{ $\pi$};
	
	\draw[right graphic] (0,1.05) node[above,align=center]{ $k$};
	\draw[right graphic] (1.05,0) node[right,align=left]{ $\omega(k)$};
	
	\draw[right graphic,very thick,blue,smooth] plot coordinates {(-1., 0.75) (-0.9, 0.734779) (-0.8, 0.693856) (-0.7, 0.636438) (-0.6, 0.570119) (-0.5, 0.5) (-0.4, 0.429881) (-0.3, 0.363562) (-0.2, 0.306144) (-0.1, 0.265221) (0., 0.25) (0.1, 0.265221) (0.2, 0.306144) (0.3, 0.363562) (0.4, 0.429881) (0.5, 0.5) (0.6, 0.570119) (0.7, 0.636438) (0.8, 0.693856) (0.9, 0.734779) (1., 0.75)};
	
	\draw[right graphic,very thick,red,smooth] plot coordinates {(-1., -0.75) (-0.9, -0.734779) (-0.8, -0.693856) (-0.7, -0.636438) (-0.6, -0.570119) (-0.5, -0.5) (-0.4, -0.429881) (-0.3, -0.363562) (-0.2, -0.306144) (-0.1, -0.265221) (0., -0.25) (0.1, -0.265221) (0.2, -0.306144) (0.3, -0.363562) (0.4, -0.429881) (0.5, -0.5) (0.6, -0.570119) (0.7, -0.636438) (0.8, -0.693856) (0.9, -0.734779) (1., -0.75)};

	\draw[right graphic,very thick, black,<->] (-0.7, 0.6) -- (-0.7, -0.6) node[pos=0.3,left]{$\ch$};
	\draw[right graphic,very thick, black,<->] (-0.67, -0.6) -- (0.67, 0.6) node[pos=0.3,left]{$\ph$};
	\draw[right graphic,very thick, black,<->] (-0.65, 0.62) -- (0.65, 0.62) node[pos=0.3,above]{$\rv$};

\end{tikzpicture} 
	\caption{Examples for band structures with different symmetries. Left: A walk with non-trivial index $\ind$. It becomes clear, that a non-trivial index is not consistent with a gap condition \cite{LongVersion}. Middle: A walk which is only chiral symmetric. The arrow denotes the action of $\ch$ on the spectrum of an admissible walk. Right: A walk with all three symmetries.}
	\label{fig:bands}
    \end{center}
\end{figure}

Many questions, e.g., the spreading of wave packets under the walk \cite{TRcoin}, can be answered by studying the eigenvalues of $\Wh(k)$ as a function of $k$. For the topological classification, however, they are completely irrelevant. Not even the details of the individual band projections $\Ba_\alpha(k)$ play a role, but only the collective projections expressing the distinction of spectra in the lower vs.\ the upper half plane:

\begin{lem}\label{lem:TIflat}
Let $W$ be a gapped translation invariant walk of any symmetry type, and let
\begin{equation}\label{upperBa}
  \Ba(k)=\sum_{\alpha,\,\omega_\alpha(k)\in(0,\pi)}\,\Ba_\alpha(k)
\end{equation}
be the band projection for the upper half plane. Then, there is a continuous path $[0,1]\ni t\mapsto W_t$ of gapped, translation invariant walks satisfying the same symmetries as $W$, such that $W_0=W$, and $\Wh_1(k)=i\Ba(k)-i(\idty-\Ba(k))=:\Wh_\flat(k)$, the {\bf flat-band walk} associated with $W$.
\end{lem}

\begin{proof}
Let $S^1_\veps$ denote the unit circle without an open disc of radius $\veps$ around each of $+1$ and $-1$. We choose $\veps$ so that the spectrum of $W$ is contained in $S^1_\veps$. Now consider a continuous function $f:[0,1]\times S^1_\veps\to S^1_\veps$, which we write as $f_t(z):=f(t,z)$, such that $f_0(z)=z$ and $f_1(z)=\sign(\im z)i$. For keeping the symmetry properties we also demand $f_t(\overline z)=\overline{f_t(z)}$. Then, we define $W_t=f_t(W)$ in the continuous functional calculus. With this all the mentioned properties are preserved, and the analogue of \eqref{Whk} reads
\begin{equation}\label{Whkt}
  \Wh_t(k)=f_t\bigl(\Wh(k)\bigr)=\sum_{\alpha=1}^d f_t\Bigl(e^{i\omega_\alpha(k)}\Bigr)\Ba_\alpha(k).
\end{equation}
\end{proof}
Clearly, for flat-band walks we have $W^2=-\idty$, i.e., $W^*=-W$. In what follows, we use this as a characterization for ``flat-band unitaries'' independently of any walk-context and the notion of bands.
Note that such continuous transformation also leaves essential locality of a walk under consideration invariant \cite{LongVersion}. For translation invariant walks with continuous band structure we show this explicitly in \Qref{pro:esslocalbands}.

Hence, for the purposes of homotopy classification it is equivalent to study either $\Wh(k)$ or the  band projections $\Ba(k)$ of $\Wh(k)$.
The symmetry conditions for these read
\begin{equation}\label{bandsym}
  \ph\Ba(k)\ph^*=\idty-\Ba(-k)    ,\ \qquad
  \rv\Ba(k)\rv^*=\Ba(-k)  ,\ \mbox{and} \qquad
  \ch\Ba(k)\ch^*=\idty-\Ba(k).
\end{equation}
These are directly equivalent to \eqref{Fousym} for the flat-band walk $W_\flat$. \Qref{fig:bands} shows typical examples of band structures together with the action of the symmetries. 
\subsection{Standard Example: The Split-Step Walk}\label{sec:splitstep}

Introduced in \cite{Kita}, this example is also treated in \cite{Kita2,Asbo1,Asbo2,Asbo4}, and many other papers. Many aspects of the current paper can be demonstrated in this example in their simplest form, which is why we introduce it early on. Some of its features can be explored with an interactive web-based tool \cite{sse}.

The coin space is two-dimensional, so a basis for $\HH$ is labelled $\ket{x,s}$ with $x\in\Ir$ and $s=\pm1$. We introduce two separate shift operations, $S_{\uparrow}$, the right shift of the spin-up vectors and $S_{\downarrow}$, the left shift of the spin-down vectors. Explicitly, $S_{\downarrow}\ket{x,-}=\ket{x-1,-}$ and $S_{\uparrow}\ket{x,+}=\ket{x+1,+}$, while leaving the opposite spins invariant, respectively. The split-step walk is then defined as
\begin{equation}\label{eq:wss}
W=BS_{\downarrow}AS_{\uparrow}B,
\end{equation}
where $B=\bigoplus_\Ir R(\theta_1/2)$ and $A=\bigoplus_\Ir R(\theta_2)$ are standard real rotation matrices acting sitewise at each $x\in\Ir$.

The symmetry of the model is of type \symBDI (see \Qref{Tab:sym}), where $\ph$ denotes the complex conjugation in position space (i.e., $\ph z \ket{x,s} = \bar{z}\ket{x,s}$ $\forall z\in\Cx$), and the chiral symmetry acts like $\sigma_1$ (i.e., $\ch\ket{x,s}=\ket{x,-s}$). The symmetry condition readily allows for non-translation invariant examples: In \eqref{eq:wss}, take $A=\bigoplus_x A_x$ and $B=\bigoplus_x B_x$ to be unitary operators, where each $A_x$, $B_x$ is admissible and acts sitewise on $\HH_x$. This is of importance when speaking of bulk-boundary correspondence, since a system in which two different phases are joined is per definition non-translation invariant.

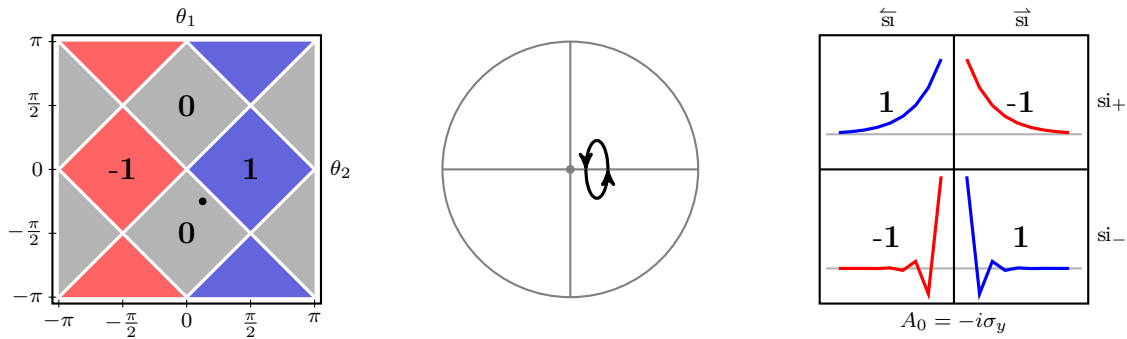
\begin{figure}
    \begin{center}
	\tikzset{
		>=stealth',
		mid graphic/.style={
			xshift=3cm
			},
		right graphic/.style={
			xshift=6cm
			},
		big arrow/.style={
			very thick,
			postaction={decorate}
		},
		big arrow reverse/.style={
			very thick,
			decoration={markings, mark=at position 0.17 with {\arrow[thick, scale=2]{>}},
				mark=at position 0.5 with {\arrow[thick,scale=2]{>}},
				mark=at position 0.83 with {\arrow[thick,scale=2]{>}}},
			postaction={decorate}
		}
	}

	\begin{tikzpicture}
	[
	scale=1.7,
	font=\footnotesize
		]
		
		\definecolor{pmcol}{RGB}{180,180,180}
		\definecolor{nncol}{RGB}{180,180,180}
		\definecolor{nmcol}{RGB}{100,100,220}
		\definecolor{pncol}{RGB}{255,100,100}

		\draw[thick] (-1.05,-1.05)  rectangle +(2.1,2.1);
		
		\fill[nncol] 	(-.5,-.5) -- +(-.5,-.5) -- +(-.5,.5)
								(-.5,-.5) -- ++(.5,.5) -- ++(.5,-.5) -- +(-.5,-.5)
								(.5,-.5) -- +(.5,.5) -- +(.5,-.5);
		\fill[pncol] 	(-.5,-.5) -- +(-.5,-.5) -- +(.5,-.5)
								(-.5,-.5) -- ++(.5,.5) -- ++(-.5,.5) -- +(-.5,-.5)
								(-.5,.5) -- +(.5,.5) -- +(-.5,.5);		
		\fill[pmcol] 	(-.5,.5) -- +(-.5,-.5) -- +(-.5,.5)
								(-.5,.5) -- ++(.5,.5) -- ++(.5,-.5) -- +(-.5,-.5)
								(.5,.5) -- +(.5,.5) -- +(.5,-.5);
		\fill[nmcol] 	(.5,-.5) -- +(-.5,-.5) -- +(.5,-.5)
								(.5,-.5) -- ++(.5,.5) -- ++(-.5,.5) -- +(-.5,-.5)
								(.5,.5) -- +(.5,.5) -- +(-.5,.5);
								
		\draw[white,very thick] (-1,-1) -- (1,1);
		\draw[white,very thick] (1,-1) -- (-1,1);
		\draw[white,very thick] (-1,0) -- (0,1) (0,-1) -- (1,0);
		\draw[white,very thick] (-1,0) -- (0,-1) (0,1) -- (1,0);
		
		\foreach \i in {-1,-.5,0,.5,1}{
			\draw[align=left] (-1.02,{\i}) -- (-1.08,{\i});
			\draw[align=left] ({\i},-1.02) -- ({\i},-1.08);
		}
		
		\draw (-1.05,-1)  node[left,align=left]{ $-\pi$};
		\draw (-1.05,-.5)  node[left,align=left]{ $-\frac{\pi}{2}$};
		\draw (-1.05,0)  node[left,align=left]{ $0$};
		\draw (-1.05,.5)  node[left,align=left]{ $\frac{\pi}{2}$};
		\draw (-1.05,1)  node[left,align=left]{ $\pi$};
		
		\draw (-1,-1.05)  node[below,align=center]{ $-\pi$};
		\draw (-.5,-1.05)  node[below,align=center]{ $-\frac{\pi}{2}$};
		\draw (0,-1.05)  node[below,align=center]{ $0$};
		\draw (.5,-1.05)  node[below,align=center]{ $\frac{\pi}{2}$};
		\draw (1,-1.05)  node[below,align=center]{ $\pi$};
		
		\fill (0.125, -.25)  circle (.03) ;

		\draw (0,1.05) node[above,align=center]{ $\theta_1$};
		\draw (1.05,0) node[right,align=left]{ $\theta_2$};

		\draw (.5,0) node[align=center]{\large$\bf{1}$};
		\draw (-.5,0) node[align=center]{\large$\bf{\text{-}1}$};
		\draw (0,-.5) node[align=center]{\large$\bf 0$};
		\draw (0,.5) node[align=center]{\large$\bf 0$};

		\draw[mid graphic,thick,gray] (0,0) circle (1);
		\draw[mid graphic,thick,gray] (-1,0) -- +(2,0);
		\draw[mid graphic,thick,gray] (0,-1) -- +(0,2);
		\draw[mid graphic, fill,gray] (0,0) circle (0.03);

		
		\draw[mid graphic, very thick, black,->] plot [smooth, tension=1] coordinates {(0.29408, 0) (0.289864, 0.0695533) (0.27763, 0.132298) (0.258574, 0.182093) (0.234563, 0.214063) (0.207946, 0.225079) (0.181329, 0.214063) (0.157318, 0.182093) (0.138262, 0.132298) (0.126028, 0.0695533) (0.121812, 0)};
		
		\draw[mid graphic, very thick, black,->] plot [smooth, tension=1] coordinates {(0.121812, 0) (0.126028, -0.0695533) (0.138262, -0.132298) (0.157318, -0.182093) (0.181329, -0.214063) (0.207946, -0.225079) (0.234563, -0.214063) (0.258574, -0.182093) (0.27763, -0.132298) (0.289864, -0.0695533) (0.29408, 0)};

		\draw[right graphic, thick] (-1.05,-1.05)  rectangle +(2.1,2.1);
		\draw[right graphic, thick] (-1.05,0) -- +(2.1,0) (0,-1.05) -- +(0,2.1);
		\draw[right graphic, thick, opacity=.3] (-1,0.275) -- (1,0.275);
		\draw[right graphic, thick, opacity=0.3] (-1,-0.775) -- (1,-.775);
		\draw[right graphic] (0,-1.05) node[below, align=center] {$A_0=-i\sigma_y$};
		\draw[right graphic] (1.05,.525) node[right,align=center] {$\six_+$};
		\draw[right graphic] (1.05,-.525) node[right,align=center] {$\six_-$};
		\draw[right graphic] (0.525,1.05) node[above,align=center] {$\sixR$};
		\draw[right graphic] (-0.525,1.05) node[above,align=center] {$\sixL$};
				
		\draw[right graphic,very thick,red] plot coordinates {(0.1, 0.863502) (0.2, 0.639821) (0.3, 0.501158) (0.4, 0.415198) (0.5, 0.361911) (0.6, 0.328877) (0.7, 0.308399) (0.8, 0.295705) (0.9, 0.287835)};	
		\draw[right graphic,very thick,blue] plot coordinates {(0.1, -0.0542976) (0.2, -0.974468) (0.3, -0.719794) (0.4, -0.790279) (0.5, -0.770771) (0.6, -0.77617) (0.7, -0.774676) (0.8, -0.77509) (0.9, -0.774975)};
		\draw[right graphic,very thick,blue] plot coordinates {(-0.9, 0.287835) (-0.8, 0.295705) (-0.7, 0.308399) (-0.6, 0.328877) (-0.5, 0.361911) (-0.4, 0.415198) (-0.3, 0.501158) (-0.2, 0.639821) (-0.1, 0.863502)};
		\draw[right graphic,very thick,red] plot coordinates {(-0.9, -0.774975) (-0.8, -0.77509) (-0.7, -0.774676) (-0.6, -0.77617) (-0.5, -0.770771) (-0.4, -0.790279) (-0.3, -0.719794) (-0.2, -0.974468) (-0.1, -0.0542976)};
			
		\draw[right graphic] (-.525,.525) node[align=center] {\large$\bf{1}$};
		\draw[right graphic] (.525,.525) node[align=center] {\large$\bf{\text{-}1}$};
		\draw[right graphic] (-.525,-.525) node[align=center] {\large$\bf{\text{-}1}$};
		\draw[right graphic] (.525,-.525) node[align=center] {\large$\bf{1}$};

	\end{tikzpicture}
	\caption{Left: Parameter plane with symmetry index $\sixR$ for the split-step walk. Middle: Winding of the upper right component of $W$ in chiral eigenbasis (see \Qref{prop:Fredformula}) for the values $(\theta_1,\theta_2)=(\pi/8,-\pi/4)$ (black dot in the left picture). Right: Eigenfunctions for the same values, after decoupling with $A_0=-i\sigma_y$. The labels in the quadrants denote the indices according to Tab. \ref{tab:indices}.}
	\label{fig:harlequin}
    \end{center}
\end{figure}

In the form of \eqref{eq:wss} it is straightforward to see that the admissibility of the $A_x$, $B_x$ suffices for the admissibility of $W$. The phase diagram for the index $\sixR(W)$ is shown in \Qref{fig:harlequin}. The main differences to the diagrams in \cite{Kita} are the signs, which is connected to the difference of methods. In \cite{Kita} phase boundaries are identified by tracking the number of bound states of compound walks. This is an intrinsically unsigned quantity, so positive and negative changes cannot be distinguished. In contrast, our approach demands also the determination of the symmetry index in the eigenspace, which is a signed integer quantity. This is crucial for the agreement with the bulk theory, in which the chiral index is a (signed) winding number.

An important feature of the index is that it allows to predict the emergence of eigenvectors at the boundary of two bulk systems in different phases (i.e. that have differing $\sixR$) that are joined (bulk-boundary correspondence). The split-step walk demonstrates this in a very simple way, by making use of \textbf{decouplings}:
To decouple the split-step walk, we replace one of the coins $A_0$ with a \textbf{splitting coin}, which forbids transition from the left to the right side and vice versa. Note that the family of \symBDI-admissible unitary operators on $\Cx^2$ consists precisely of the rotation matrices $R(\theta)$ and $\pm\sigma_1$, which are in different connected components. This allows for \emph{non-gentle} decouplings, which leave $\six$ and $\sixR$ invariant, but change $\sixR_+$ and $\sixR_-$ \cite{LongVersion}.
Only the splitting coin chosen from the rotations, $A_0=\pm R(\pi/2)=\pm i \sigma_2$ is a gentle perturbation, which implies that eigenvalues at $+1$ (and $-1$) come in pairs with opposing chirality (see \Qref{fig:harlequin}).

\subsection{Dependence on cell structure} \label{sec:stabelHomo}
The homotopy classification we are aiming at is an equivalence relation of the type ``$W_1$ can be transformed into $W_2$''. It clearly depends very sensitively to what kind of transformations are allowed. Allowing additional operations coarsens the equivalence relation. Therefore, our strategy in the following will be {\it not} to allow these operations at first. This leads to a {\it finer} classification of translation invariant walks than our general theory. Apart from continuous deformations there are three operations we need to consider, all of which are related to reorganizing the cell structure. Our general theory is naturally insensitive to the first two, and has a predictable behaviour under the third.

\subsubsection{Regrouping}\label{sec:regrouping}

For the general theory outlined in \Qref{sec:nonti} the details of cell structure are irrelevant: the local Hilbert spaces may have different dimensions, and nothing changes, if we decide to consider two neighbouring cells together as a new single cell. Translation invariance makes sense only if the cells all have the same dimension. Similarly, only regrouping operations done uniformly throughout the lattice can be considered, like always grouping cell $\HH_{2x}$ together with cell $\HH_{2x+1}$. The main change introduced by this operation is a change in the definition of the translations: After regrouping effectively only translations by an even number of sites are considered. This makes a big difference, if we demand that translation invariance be conserved on a deformation path. Specifically, it might be impossible to deform $W_1$ into $W_2$ keeping translation invariance, but it might be possible to do so by breaking the translation invariance and keeping only the weaker requirement of period-2 translation invariance. We will see examples of this below.

For later use we record $\Wh$ for a regrouped walk with the simple pair-regrouping indicated above. For the regrouped walk we have the double cell dimension, and vectors  $\psi\in\HH$ get split into
\begin{equation}\label{psiregroup}
  \psi_r(x)=\begin{pmatrix}\psi_e(x)\\ \psi_o(x)\end{pmatrix}=\begin{pmatrix}\psi(2x)\\ \psi(2x+1)\end{pmatrix}.
\end{equation}
The regrouped walk $W_r$ acts in position space by convolution with
\begin{equation}\label{Wregroup}
   W_r(x)=\begin{pmatrix}W(2x)&W(2x-1)\\W(2x+1)&W(2x)\end{pmatrix},
\end{equation}
and by taking the Fourier transform we get
\begin{equation}\label{Whregroup}
\Wh_r(k)= H(k/2)\begin{pmatrix}\Wh\bigl(\frac k2\bigr)&0\\0&\Wh\bigl(\frac k2+\pi\bigr)\end{pmatrix}H(k/2)^*
         , \qquad \text{where}\qquad
         H(k)=\frac{1}{\sqrt2}\begin{pmatrix}1&1\\e^{-ik}&-e^{-ik}\end{pmatrix}.
\end{equation}

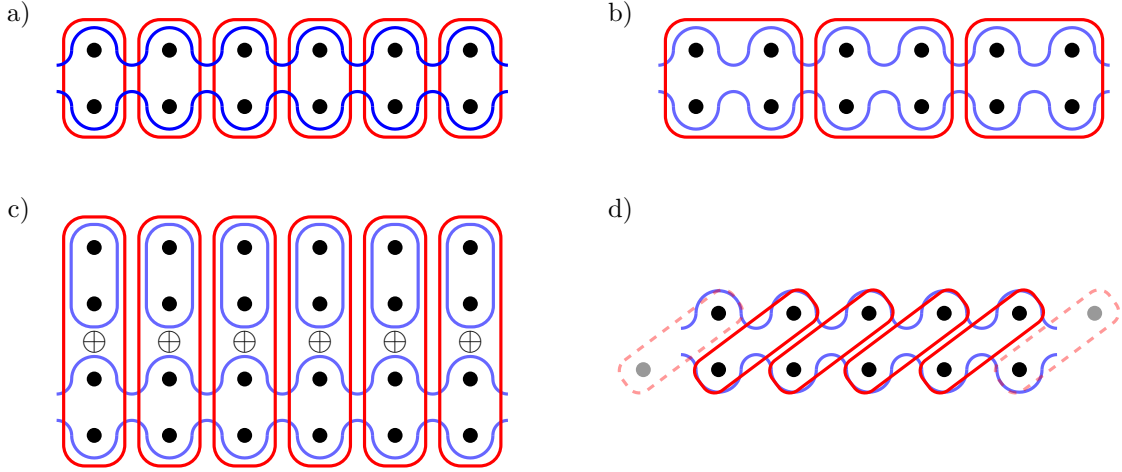
\begin{figure}
    \begin{center}
%
%
\def\xdist{4}
\def\ydist{1.75}
\def\xwid{7}
\def\ywid{3}
\def\rs{.3}
\def\rb{.2}
\def\opa{0.6}
\tikzstyle{dot} =[circle,fill,inner sep = 2]

\begin{tikzpicture}

\node[inner sep=0pt] (walk) at (-\xdist,\ydist)
{

\begin{tikzpicture}
	\foreach \i in {-2,...,2,3}{
		\node[dot] (u\i) at (\i,0.75) {};
		\node[dot] (d\i) at (\i,0) {};
		\draw [blue,very thick,domain=0:180] plot ({\i+\rs*cos(\x)}, {.75+\rs*sin(\x)});
		\draw [blue,very thick,domain=180:360] plot ({\i+\rs*cos(\x)}, {\rs*sin(\x)});
		\node[rectangle,rounded corners=8,very thick,draw=red,fit=(u\i) (d\i),inner sep=3mm](te\i) {};

	};
	\foreach \i in {-2,...,2}{
		\draw [blue,very thick,domain=180:360] plot ({\i+0.5+\rb*cos(\x)}, {.75+\rb*sin(\x)});
		\draw [blue,very thick,domain=0:180] plot ({\i+0.5+\rb*cos(\x)}, {\rb*sin(\x)});
	};
	\draw [blue,very thick,domain=180:270] plot ({3.5+\rb*cos(\x)}, {.75+\rb*sin(\x)});
	\draw [blue,very thick,domain=180:90] plot ({3.5+\rb*cos(\x)}, {\rb*sin(\x)});
	
	\draw [blue,very thick,domain=360:270] plot ({-2.5+\rb*cos(\x)}, {.75+\rb*sin(\x)});
	\draw [blue,very thick,domain=0:90] plot ({-2.5+\rb*cos(\x)}, {\rb*sin(\x)});
		
\end{tikzpicture}};
\node (A) at (-7.5,2.6)  {a)};

\node[inner sep=0pt] (regroup) at (\xdist,\ydist)
{\begin{tikzpicture}
\foreach \i in {-2,...,2,3}{
	\node[dot] (u\i) at (\i,0.75) {};
	\node[dot] (d\i) at (\i,0) {};\
	\draw [opacity=\opa,blue,very thick,domain=0:180] plot ({\i+\rs*cos(\x)}, {.75+\rs*sin(\x)});
	\draw [opacity=\opa,blue,very thick,domain=180:360] plot ({\i+\rs*cos(\x)}, {\rs*sin(\x)});
	};

	\foreach \i in {-2,...,2}{
	\draw [opacity=\opa,blue,very thick,domain=180:360] plot ({\i+0.5+\rb*cos(\x)}, {.75+\rb*sin(\x)});
	\draw [opacity=\opa,blue,very thick,domain=0:180] plot ({\i+0.5+\rb*cos(\x)}, {\rb*sin(\x)});
};

	\draw [opacity=\opa,blue,very thick,domain=180:270] plot ({3.5+\rb*cos(\x)}, {.75+\rb*sin(\x)});
	\draw [opacity=\opa,blue,very thick,domain=180:90] plot ({3.5+\rb*cos(\x)}, {\rb*sin(\x)});

	\draw [opacity=\opa,blue,very thick,domain=360:270] plot ({-2.5+\rb*cos(\x)}, {.75+\rb*sin(\x)});
	\draw [opacity=\opa,blue,very thick,domain=0:90] plot ({-2.5+\rb*cos(\x)}, {\rb*sin(\x)});

	\node[rectangle,rounded corners=8,very thick,draw=red,fit=(u-2) (d-2) (u-1) (d-1),inner sep=3mm](re1) {};
	\node[rectangle,rounded corners=8,very thick,draw=red,fit=(u0) (d0) (u1) (d1),inner sep=3mm](re1) {};
	\node[rectangle,rounded corners=8,very thick,draw=red,fit=(u2) (d2) (u3) (d3),inner sep=3mm](re1) {};
\end{tikzpicture}};
\node (B) at (.5,2.6) {b)};

\node[inner sep=0pt] (add) at (-\xdist,-\ydist)
{\begin{tikzpicture}

	\foreach \i in {-2,...,2,3}{
		\node[dot] (u\i) at (\i,0.75) {};
		\node[dot] (d\i) at (\i,0) {};\
		\draw [opacity=\opa,blue,very thick,domain=0:180] plot ({\i+\rs*cos(\x)}, {.75+\rs*sin(\x)});
		\draw [opacity=\opa,blue,very thick,domain=180:360] plot ({\i+\rs*cos(\x)}, {\rs*sin(\x)});
	};

	\foreach \i in {-2,...,2}{
		\draw [opacity=\opa,blue,very thick,domain=180:360] plot ({\i+0.5+\rb*cos(\x)}, {.75+\rb*sin(\x)});
		\draw [opacity=\opa,blue,very thick,domain=0:180] plot ({\i+0.5+\rb*cos(\x)}, {\rb*sin(\x)});
	};

	\draw [opacity=\opa,blue,very thick,domain=180:270] plot ({3.5+\rb*cos(\x)}, {.75+\rb*sin(\x)});
	\draw [opacity=\opa,blue,very thick,domain=180:90] plot ({3.5+\rb*cos(\x)}, {\rb*sin(\x)});

	\draw [opacity=\opa,blue,very thick,domain=360:270] plot ({-2.5+\rb*cos(\x)}, {.75+\rb*sin(\x)});
	\draw [opacity=\opa,blue,very thick,domain=0:90] plot ({-2.5+\rb*cos(\x)}, {\rb*sin(\x)});

	\foreach \i in {-2,...,2,3}{
		\node[dot] (tu\i) at (\i,2.5) {};
		\node[dot] (td\i) at (\i,1.75) {};\
		\node[rectangle,rounded corners=8,very thick, opacity=\opa,draw=blue,fit=(tu\i) (td\i),inner sep=2mm](te\i) {};
		\node[rectangle,rounded corners=8,very thick,draw=red,fit=(u\i) (d\i) (tu\i) (td\i),inner sep=3mm](te\i) {};
		\node[] (p\i) at (\i,1.25) {\scalebox{1.25}{$\oplus$}};
	};
\end{tikzpicture}};
\node (C) at (-7.5,0) {c)};

\node[inner sep=0pt] (slice) at (\xdist-.2,-\ydist)
{\begin{tikzpicture}
	\foreach \i in {-2,...,2}{
		\node[dot] (u\i) at (\i,0.75) {};
		\node[dot] (d\i) at (\i,0) {};\
		\draw [opacity=\opa,blue,very thick,domain=0:180] plot ({\i+\rs*cos(\x)}, {.75+\rs*sin(\x)});
		\draw [opacity=\opa,blue,very thick,domain=180:360] plot ({\i+\rs*cos(\x)}, {\rs*sin(\x)});
	};

	\foreach \i in {-2,...,1}{
		\draw [opacity=\opa,blue,very thick,domain=180:360] plot 	({\i+0.5+\rb*cos(\x)}, {.75+\rb*sin(\x)});
		\draw [opacity=\opa,blue,very thick,domain=0:180] plot ({\i+0.5+\rb*cos(\x)}, {\rb*sin(\x)});
	};

	\draw [opacity=\opa,blue,very thick,domain=180:270] plot ({2.5+\rb*cos(\x)}, {.75+\rb*sin(\x)});
	\draw [opacity=\opa,blue,very thick,domain=180:90] plot ({2.5+\rb*cos(\x)}, {\rb*sin(\x)});

	\draw [opacity=\opa,blue,very thick,domain=360:270] plot ({-2.5+\rb*cos(\x)}, {.75+\rb*sin(\x)});
	\draw [opacity=\opa,blue,very thick,domain=0:90] plot ({-2.5+\rb*cos(\x)}, {\rb*sin(\x)});

	\draw[rounded corners=5,rotate around={-53.1:(0,0)},very thick,red] (-.25,-.3) rectangle (.25,1.55);
	\draw[rounded corners=5,rotate around={-53.1:(-2,0)},very thick,red] (-2.25,-.3) rectangle (-1.75,1.55);
	\draw[rounded corners=5,rotate around={-53.1:(-1,0)},very thick,red] (-1.25,-.3) rectangle (-0.75,1.55);
	\draw[rounded corners=5,rotate around={-53.1:(1,0)},very thick,red] (0.75,-.3) rectangle (1.25,1.55);

	\draw[rounded corners=5,rotate around={-53.1:(2,0)},very thick,red,opacity=.4,dashed] (1.75,-.3) rectangle (2.25,1.55);
	\draw[rounded corners=5,rotate around={-53.1:(-3,0)},very thick,red, opacity=.4,dashed] (-3.25,-.3) rectangle (-2.75,1.55);
	
	\node[dot,opacity=0.4] (d-3) at (-3,0) {};
	\node[dot,opacity=0.4] (u4) at (3,.75) {};
\end{tikzpicture}};
\node (D) at (.5,0)  {d)} ;

 \end{tikzpicture}

	\caption{Visualization of the different transformations of the cell structure discussed in \Qref{sec:stabelHomo}. The red boxes indicate the different choices of unit cells: a) original cell structure, b) pairwise regrouped cells, c) addition of a trivial walk, d) skew cut cells.}
	\label{fig:regroupeddomino}
    \end{center}
\end{figure}

\subsubsection{Adding trivial walks}
A second operation which one might consider for transforming $W_1$ into $W_2$ is adding trivial systems, which have to be returned unchanged after the transformation. A walk $W^0$ is said to be ``trivial'' in this context if it acts cell-wise, i.e.\  there is no propagation between cells. Note that this condition immediately implies $\sixR(W^0)=0$. In this case, although $W_1$ could not be transformed into $W_2$, it might be possible that ``there are trivial walks $W_1^0$ and $W_2^0$ so that $W_1\oplus W_1^0$ can be continuously deformed into $W_2\oplus W_2^0$''. It is one of the basic features of our general theory that indices add up for direct sums of walks, and also that the value of the index lies in a group. Together, this implies a cancellation law, so that allowing the addition of trivial walks does not change the general classification. Note that since this classification is a complete homotopy invariant, walks that can be deformed into each other after adding trivial ones can actually be deformed into each other when a breaking of translation invariance is allowed on the way.
Later, in the discussion of completeness, we will show that such trivial walks indeed always exist.

There is a general construction for turning an invariant which is additive over direct sums (with values in a semigroup) into a group-valued invariant with cancellation law. This is called the Grothendiek group of the semigroup, and is a standard ingredient of K-theory.

\subsubsection{Time frames and skew recutting of cells}
Suppose a walk can be written as $W=W_1W_2$, i.e., there are two kinds of steps which alternate. Then there is a closely related walk $W'=W_2W_1$, alternating the same two steps, but beginning with the other one. In this sense the two walks differ by a choice of ``time frame'' \cite{Asbo2}. For the long time behaviour there should be little difference, and indeed the spectra are the same, because $W'=W_1^*WW_1$. Hence, $\six_\pm$ are left invariant. But since $\sixR$ is sensitive to the cell structure $W$ is defined on, it might change  under such operation, if $W_1$ does not act site-wise, i.\ e.\ $[W_1,P_{\geq 0}]\neq 0$.
Consider for example the split-step walk, with $W_1=BS_\downarrow\sqrt{A}$ and $W_2=\sqrt{A}S_\uparrow B$. $W'$ is of the same form as $W$, with $\theta_1$ and $\theta_2$ as well as $S_\uparrow$ and $S_\downarrow$ swapped. This results in a $\pi/2$-clockwise rotation of the index-plane in \Qref{fig:harlequin} and hence, trivial phases become non-trivial and vice versa. The dependence on the cell structure is also easily obtained from the example in \Qref{fig:regroupeddomino} (bottom-right). Assume the original walk to act locally in each cell, i.e., $\ker(PWP)=0$ on $P\HH$. Then the new cell structure produces a non-trivial kernel of $PW'P$ on $P\HH$ and hence, $\sixR$ might change.

\section{Essential locality and continuity of band structure}\label{sec:essloc}

As for all Fourier transforms, decay properties of $W(x)$ translate into smoothness conditions on $\Wh$. We refer to such conditions as locality conditions, since they express the idea that in a single step the walker cannot go very far, or at least, that the amplitude for very long jumps goes down sufficiently fast. Some condition of this kind is needed to make the theory non-trivial: Without a localization requirement along the path any two unitaries could be continuously connected (see \Qref{sec:ContractShift} for an example of the kind of locality violation this may imply). In order to classify as many systems as possible, one would like to assume as little smoothness/locality as possible. For the classification of band structures the natural smoothness condition appears to be continuity of $\Wh$. Indeed, the formulas discussed below are often in terms of winding numbers, which are well-defined assuming just continuity. On the other hand, it is easy to see that with admission of discontinuities no classification is possible. We show here (\Qref{pro:esslocalbands}) that in the translation invariant setting this is the same as saying that the walk operator can be uniformly approximated by walks with strictly finite jump length. This will be our standing assumption in the sequel.

Of course, we verify also that this condition implies ``essential locality'' as used in the the general theory \cite{LongVersion}. Surprisingly, however, it turns out that the converse is not true: There are essentially local, translation invariant walks such that $\Wh$ is {\it not} continuous. Nevertheless these are covered by the general classification, so we conclude that winding numbers can also be extended as good topological classifiers beyond continuous curves! This generalization has been noted before, leading to the notion of ``quasi-continuous'' \cite{Peller} functions on the circle. We discuss this in \Qref{sec:quasiCont}. While this distinction may be a subtlety of little practical value, it needs to be taken seriously, when one wants to go to higher lattice dimensions. It is not completely clear what locality conditions are appropriate then, and lead to a manageable theory. Basically, one needs to fix a ``coarse structure'' \cite{RoeIndex,RoeCoarse}, and for this, too, the one-dimensional case already has two natural options: These are ``approximability by walks with uniformly finite jump length'' and the ``essential locality'', which in the translation invariant case map to continuity and quasi-continuity of $\Wh$, respectively.

\subsection{Approximation by strictly local operators}
We begin with a general result which relates smoothness conditions on $\Wh$ to locality properties of the corresponding $W$:

\begin{prop}\label{pro:esslocalbands} Let $A\in\BB(\HH)$, $\HH=\ell^2(\Ir)\otimes\HH_0$ be translation invariant with Fourier transform $\Ah$.
Then the following are equivalent:
\begin{itemize}
	\item[(1)] $A$ can be approximated in norm by strictly local operators.
	\item[(2)] $A$ can be approximated in norm by translation invariant operators $A_n$ with $\Ah_n \in {\cal C}^\infty$.
	\item[(3)] $k\mapsto\Ah(k)$ is a continuous function with periodic boundary conditions.
\end{itemize}
If $A$ is unitary, the approximating operators in  (2) can be chosen to be unitary and admissible for all symmetries under consideration.
Moreover, if one of the above conditions is fulfilled, $A$ is essentially local, i.e.\ $[P,A]$ is compact, where $P=\Pg0$ denotes the projection onto the positive positions.
\end{prop}

\begin{proof}
	(3)$\Rightarrow$(1):\ This is basically the Stone-Weierstra\ss\ Theorem \cite{StoneWeier,weier}, which immediately implies that the algebra of trigonometric polynomials is sup-norm dense in the set of continuous functions.
	Detailed error estimates are provided by the theory of Fourier series \cite[Sect.\;1.4]{Dym}. There it is shown that, for any continuous function $\widehat A(k)$ on the circle, a suitable sequence of truncations of the Fourier series, for example, those given by the Fej\'er kernels $\widehat A_n(k)$, converge uniformly to the given function. The same estimates work also for matrix valued functions, providing us with a sequence $A_n$ of strictly local operators such that $\norm{A_n-A}\leq\varepsilon_n\to0$.
	
	(1)$\Rightarrow$(2, unitarity)$\Rightarrow$(3):\ Since the Fourier transform of a strictly local operator is a trigonometric polynomial, (1) obviously implies (2). While the choice of the Ces\`aro approximants $\widehat{A}_n$ given by the Fej\'er kernels guarantees that $A_n$ will satisfy the same symmetries as $A$, and will be hermitian if $A$ is, unitarity is not preserved, and it is unclear whether one can modify $A_n$ to be strictly unitary. On the other hand,  $\sup_k\norm{\Ah_n(k)^*\Ah_n(k)-\idty}=\norm{A_n^*A_n-\idty}\leq2\varepsilon_n$, so the inverse square root $\widehat M_n(k)=(\Ah_n(k)^*\Ah_n(k))^{-1/2}$ is infinitely differentiable, since it can be obtained by applying a convergent power series to a polynomial. Moreover, $\widehat U_n(k)=\Ah_n(k)\widehat M_n(k)$ is unitary and converges in norm to $A$. Being the polar isometry of an admissible operator, $\widehat U_n(k)$ is also admissible.
	Admissibility for $\ph$ follows trivially, since $\widehat M_n(k)$ is a real function of an admissible operator. For any of the conjugating symmetries $\sigma\in\{\gamma,\tau\}$ we get
	\begin{equation}
	\sigma U=\sigma A(A^*A)^{-1/2}=A^*(AA^*)^{-1/2}\sigma=((AA^*)^{-1/2}A)^*\sigma=U^*\sigma.
	\end{equation}	
	For any $A$ allowing such approximation it follows that $\Ah$ is continuous as the norm limit of continuous functions.
	Note, that convergence of $A_n\rightarrow A$ in norm also implies convergence of the spectra. Therefore, any gap of $A$ eventually becomes a gap for $A_n$ for large enough $n$.
	
	Now suppose $A$ is strictly local. Consider the commutator $[P,A]=PA(\idty-P)-(\idty-P)AP$ as a block matrix with respect to the cell decomposition $\HH=\bigoplus\HH_x$. It contains only terms between the subspaces $\HH_x$ and $\HH_y$, where $x\geq0$ and $y<0$, or conversely $x<0$ and $y\geq0$. Moreover, since $A$ is local, there are only non-zero blocks with $\abs{x-y}\leq N$. Since there are only finitely many such pairs $(x,y)$ we conclude that $[P,A]$ is a finite rank operator. When $A$ is approximated in norm by such operators, $[P,A]$ is approximated in norm by finite rank operators, hence compact.
\end{proof}

For the homotopy classification of translation invariant quantum walks we will restrict ourselves to walks, which fulfil the conditions of \Qref{pro:esslocalbands}. This guarantees that we never leave the set of essentially local walks and as we show below the classification of such systems already turns out to be complete.

\subsection{Quasi-continuity}\label{sec:quasiCont}
Let us, however, state what the condition of essential locality exactly translates to in terms of the band structure. We need to express that $[P,W]$ is a compact operator, which is equivalent to the compactness of the two operators $PW(\idty-P)$ and $(\idty-P)WP$. These are the off-diagonal quadrants of $W$ in a representation of $W$ as a doubly infinite matrix of $d\times d$ blocks. By translation invariance the $(x,y)$-block is $W(x-y)$.

Relabelling $y\mapsto y'=-y$ for the quadrant with indices $y<0\leq x$ leads to a block Hankel matrix $M$, i.e., a matrix whose block entries $M(x,y')$ are labelled by $x\geq0$, $y'>0$, and $M(x,y')=m(x+y')$ depends only on $(x+y')$. The characterization of compact Hankel matrices with scalar entries is achieved by a classic result known as Hartman's Theorem \cite{Hartman}. It says that the numbers $m(x)$ are Fourier coefficients of a continuous function. Hartmann's Theorem generalizes to block matrices. This is called the ``vectorial'' case in \cite[Ch.2]{Peller}, where the relevant result is Thm.~4.1, and allows even for block entries that are operators between distinct, possibly infinite Hilbert spaces. The conclusion is entirely the same, i.e., we get that there is a continuous matrix valued function $k\mapsto\widehat M$ on $[-\pi,\pi]$ with periodic boundary condition, whose Fourier coefficients for $x\geq0$ equal $W(x)$.

The notion of quasi-continuity arises from the peculiar way in which we have to combine two such conditions. The function spaces involved are the matrix valued versions of $L^\infty$, the Banach algebra of essentially bounded measurable functions on the unit circle,  its  subalgebra $C$ of continuous functions, and $H^\infty$, the subalgebra with vanishing Fourier coefficients for negative indices. A (possibly matrix valued) function $f$ is called \textbf{quasi-continuous}, if $f=g+h$, with $g\in C$ and $h\in H^\infty$, and such a decomposition also holds for $f^*$ \cite{Peller}. Indeed, this is the required property:
We have, on the one hand, $\Wh=\widehat M+(\Wh-\widehat M)\in C+\overline{H^\infty}$, because only the negative Fourier coefficients can be non-zero. On the other, evaluating the compactness condition for the second block, we get the analogous condition for $W(x)$ with $x<0$, which means that $\Wh^*\in C+\overline{H^\infty}$. Hence $\Wh$, or equivalently each of its entries is quasi-continuous.

We summarize this discussion in the following proposition.

\begin{prop}\label{prop:quasiConti}
	Let $A\in\BB(\HH)$, with $\HH=\ell^2(\Ir)\otimes\HH_0$, $\HH_0$ finite dimensional, be translation invariant and $\Ah$ its Fourier transform.
	Then the following are equivalent:
	\begin{itemize}
		\item[(1)] $A$ is essentially local.
		\item[(2)] $\Ah$ is quasi-continuous.
	\end{itemize}
\end{prop}

Of course, it is a basic fact of this theory that there are quasi-continuous functions which are not continuous. For a rough sketch how counterexamples arise, note that, for a continuous function $f\in C$, $f(k)$ traces out a curve in the complex plane, and the $H^\infty$ condition makes this the boundary of the image of the unit disk under an analytic function. On the other hand, the Riemann mapping theorem also works for open sets which are not bounded by a continuous curve. So discontinuous examples are constructed in terms of the boundary values of analytic functions mapping the open unit disk to some such wild region. To build an essentially local walk with discontinuous $\Wh$ we disregard the symmetry conditions for the moment, and take $\dim\HH_0=1$, i.e., the scalar case. Then we set $\Wh(k)=\exp\bigl(i\im f(e^{ik})\bigr)$, where $f$ is a conformal mapping with continuous real part, but only quasi-continuous imaginary part (see  \cite[p.377]{garnett2007bounded} for an explicit example).

Winding numbers still make sense as the limit of the winding numbers of curves just to the inside of the boundary \cite{Sarason,Peller}. They also coincide with the (negative) Fredholm index of $PWP$ \cite{Douglas}, i.e. with the index from \Qref{def:oldindex}. In the above example we get $\ind W=0$, because
$\Wh(k) = \exp\bigl(-\re f(e^{ik})\bigr)\exp\bigl(f(e^{ik})\bigr)$. Here the first factor is positive and hence does not contribute to any winding. The second is the boundary value of a function which is analytic in the open disk, so its generalized winding number, evaluated close to the boundary, vanishes.

\subsection{Examples}\label{sec:essloc_examples}
\subsubsection{Exponentials}
Consider an essentially local Hamiltonian $H$, and define $W=\exp(iHt)$, for $t\in\Rl$. Then $W$ is also essentially local, because $[P,H]$ compact means that the images of $P$ and $H$ commute in the Calkin algebra (bounded operators modulo compact ones), which implies the same for $W$. Another way to put the statement is that the essentially local operators form a C*-algebra, so functions of the (multivariate) continuous functional calculus will always map essentially local arguments to essentially local operators.

For translation invariant operators with continuous bands this implication is even simpler, namely that the continuity of $k\mapsto \widehat H(k)$ implies the continuity of $k\mapsto\exp(i\widehat H(k)t)$. Note, however, that no such conclusion is available for strictly local Hamiltonians and their exponentials.

\subsubsection{Contracting the shift}\label{sec:ContractShift}
The bilateral shift on $\ell^2(\Ir)$ is the fundamental example of a strictly local, translation invariant walk, which cannot be contracted to the identity, while keeping (strict) locality.
In fact, the index \eqref{eq:indW} of one-dimensional walks 
can be understood as the ``shift content'' of $W$, and at the same time labels the homotopy classes of walks for deformations keeping strict locality. Here we revisit this example and show that one can also not contract the shift to the identity under the assumption of mere essential locality. This follows also from the extension of index theory in \cite[]{LongVersion}, which establishes the index as a Fredholm index. But how badly does essential locality fail on a deformation path?

The Fourier matrix of the shift $\ket x\mapsto\ket{x+1}$
is given by $\Wh(k)=\exp(ik)$. A natural attempt to contract this to the identity, i.e., to make ``fractional steps'' on the lattice, is by the operator family $W_\lambda$ with
\begin{equation}\label{Whhalfshift}
  \Wh_\lambda(k)=e^{i\lambda k}\idty,
\end{equation}
where $0<\lambda<1$. These interpolate between the identity ($\lambda=0$) and the shift ($\lambda=1$) in a norm-continuous way.
Despite appearances this is {\it not} a continuous function of $k$, since for this we also require periodic boundary conditions (or else our result would depend on the arbitrary choice of fundamental domain $[k_0,k_0+2\pi]$). Indeed \eqref{Whhalfshift} jumps from $\exp(-i\lambda\pi)$ to $\exp(+i\lambda\pi)$, as $k$ crosses from $-\pi$  to $\pi$.  However, $W_\lambda$ is also not essentially local for any $\lambda$.

To see this, take the inverse Fourier transform to get the spatial convolution kernel
\begin{equation}\label{halfshift1}
  W_\lambda(x)= \frac{\sin(\pi\lambda)}\pi\,\frac{(-1)^x}{\lambda-x}.
\end{equation}
We have to decide whether $PW_\lambda(\idty-P)$ is compact. In order to bring it into a more familiar form we introduce the operator $S:P\HH\to(\idty-P)\HH$ given by $(S\psi)(x)=\psi(-1-x)$, and find
\begin{equation}\label{halfshift2}
 \bigl(PW_\lambda(\idty-P)S\psi\bigr)(x)=\frac{\sin(\pi\lambda)}\pi\,\sum_{y\geq0}\frac{(-1)^{x+y}}{x+y+1-\lambda}\psi(y).
\end{equation}
Up to the prefactor and the conjugation by the unitary operator multiplying with alternating signs, this is a ``generalized Hilbert matrix'', which has been studied in detail in the 50s.
It was found \cite[Thm.~5]{Rosenblum} that it has a continuous spectral component for all $\lambda\in(0,1)$, and hence cannot be compact. So indeed $W_\lambda$ is not essentially local for any $0<\lambda<1$.

\subsubsection{Decay properties of walk matrix elements}\label{sec:decay_mat_elements}
We can write a general walk as $(W\psi)(x)=\sum_yW(x,y)\psi(y)$, where $W(x,y):\HH_y\to\HH_x$, and wave functions are written in terms of their local components $\psi=\bigoplus_x\psi(x)$.
Strict locality means that $W(x,y)=0$ for $\abs{x-y}>L$, for some $L$. More generally one could look at decay properties like
\begin{equation}\label{WxyDecay}
  \norm{W(x,y)}\leq c\,\abs{x-y}^{-\alpha},
\end{equation}
whenever $x\neq y$. In the translation invariant case, this can be specialized by setting $W(x,y)=W(x-y)$.

A simple sufficient criterion is obtained by computing the Hilbert-Schmidt norm of $PW(\idty-P)$. If this is finite, this operator is compact, and a similar criterion for $(\idty-P)WP$ gives the compactness of
$[P,W]=PW(\idty-P)-(\idty-P)WP$. We get
\begin{eqnarray}\label{Wdecay1}
 \norm{PW(\idty-P)}_2^2
    &=& \sum_{x\geq0>y} \tr  W(x,y)^*W(x,y)
    \leq  d \sum_{x\geq0>y} \norm{ W(x,y)}^2\\
    &\leq& cd \sum_{x\geq0>y} \abs{y-x}^{-2\alpha}
    =  cd\sum_{n=1}^\infty n\cdot n^{-2\alpha}, \label{Wdecay2}
\end{eqnarray}
where in \eqref{Wdecay1} we used that $\norm A_2^2\leq d\norm A^2$ for a complex $d\times d$-matrix $A$, and in \eqref{Wdecay2} that there are exactly $n$ terms with $x-y=n$ in the first sum. The estimate converges if
$1-2\alpha<-1$, i.e., $\alpha>1$. This is exactly the condition we get in the translation invariant case for $\sum_x\norm{W(x)}<\infty$, which implies uniform convergence of $\Wh_N(k)=\sum_{\abs{x}\leq N}W(x)e^{i k x}$ to $\Wh(k)$ and, therefore, by the uniform convergence theorem will imply the continuity of $\Wh$.

To summarize, $\alpha>1$ in \eqref{WxyDecay} is sufficient for essential locality. This is optimal, because
the interpolated shift gives a counterexample with $\alpha=1$.

\section{Homotopy classification for translation invariant systems}\label{sec:classification}

In our setting, we distinguish two different kinds of symmetry types with non-trivial index group: those with index group isomorphic to $\Ir$ and those with index groups isomorphic to $\Ir_2$. Bearing in mind the convention of trivial multiplication phases of the symmetries, the symmetry types with index group $\Ir$ are distinguished by the presence of a chiral symmetry which squares to $+\idty$, as can be read off from Table \ref{Tab:sym}. The index group $\Ir_2$ is obtained for the symmetry types where there is no chiral symmetry or where it squares to $-\idty$.

In this section we will first examine the only symmetry type without chiral symmetry, i.e. symmetry type \symD. In a later section we consider the symmetry types which include a chiral symmetry, where \symDIII plays a special role. Although having a $\Ir_2$-valued index group we treat it together with the symmetry types with $\ch^2=+\idty$, since the chiral symmetry strongly influences the structure of symmetric unitaries also in that case.

In both cases we obtain a homotopy classification which we prove to be complete in the sense that translation invariant walks with equal indices are deformable into each other along an admissible path of translation invariant walks. However, we caution the reader that for $\symS=\symD,\symBDI$ additional invariants appear if we insist on a given cell structure. Regrouping of the cells or adding trivial systems allows us to trivialize these additional invariants. Additionally, we prove concrete index formulas for all symmetry types.

As already mentioned in the introduction, there is a continuous mapping between translation invariant quantum walks and ``effective Hamiltonians''. The homotopy classification is therefore the same in both settings, and in particular the completeness results derived in this section carry over to effective Hamiltonians. 
\subsection{Particle-hole symmetric walks}\label{sec:Dwalks}

Consider the case of symmetry type \symD, with only the particle-hole symmetry $\ph$ and $\ph^2=\idty$. In this case the index is just a parity. By \Qref{lem:TIflat} we can restrict consideration to flat-band walks, which are equivalently given by the continuous family of band projections $\Ba(k)$ satisfying
\begin{equation}\label{bandsymPH}
\ph\Ba(k)\ph=\idty-\Ba(-k).
\end{equation}
By this equation we only need to specify $\Ba(k)$ for $k\in[0,\pi]$, and \eqref{bandsymPH} imposes no constraint on $\Ba(k)$ for $0<k<\pi$. However, at the end points we get projections with a special property.

\def\etflipp{{\mathcal P}_\ph}

\begin{defi}\label{def:etaflipped}
	Let $\ph$ be an antiunitary operator on a finite dimensional Hilbert space $\HH$ with $\ph^2=\idty$.
	Then we call a projection $\Ba$ {\bf $\ph$-flipped} if
	$\ph \Ba\ph+\Ba=\idty$. The set of such projections will be denoted by $\etflipp$.
\end{defi}

Since $\Ba$ and $\ph \Ba\ph$ obviously have the same dimension, $\HH$ must be even dimensional, and we write $\dim\HH=2d$.

The basic observation about $\etflipp$ is that it has two connected components, described in the following Lemma.

\begin{lem}\label{lem:etaflip}
	In the setting of \Qref{def:etaflipped}  let $\Ba,\Ba'\in\etflipp$.
	Introduce orthonormal systems $\phi_\alpha\in \Ba\HH$ and  $\phi'_\alpha\in \Ba'\HH$ ($\alpha=1,\ldots,d$) spanning these spaces, and extend them to bases of $\HH$ by setting $\phi_{\alpha+d}=\ph\phi_\alpha$ and $\phi'_{\alpha+d}=\ph\phi'_\alpha$.
	Let $M_{\alpha\beta}=\braket{\phi_\alpha}{\phi'_\beta}$ for $\alpha,\beta=1,\ldots,2d$, and consider $s(\Ba,\Ba')=\det M$. Then
	\begin{itemize}
		\item[(1)] When $\Ba'=N\Ba N^*$ with $N$ unitary and $N\ph=\ph N$, $s(\Ba,\Ba')=\det N$.
		\item[(2)] $s(\Ba,\Ba')s(\Ba',\Ba'')=s(\Ba,\Ba'')$.
		\item[(3)] $s(\Ba,\Ba')$ is independent of the choice of $\phi_\alpha\in \Ba\HH$ and $\phi'_\alpha\in \Ba'\HH$.
		\item[(4)] $s(\Ba,\Ba')=\pm1$.
		\item[(5)] $s(\Ba,\Ba')$ depends continuously on $\Ba$ and $\Ba'$.
		\item[(6)] $s(\Ba,\Ba')=1$ if and only if $\Ba$ and $\Ba'$ can be connected continuously inside $\etflipp$.
	\end{itemize}
\end{lem}

\begin{proof}
	Until (3) is established, consider $s(\Ba,\Ba')$ as a quantity which depends not just on $\Ba$ and $\Ba'$, but also on the bases chosen. Then (1) is basically a reformulation of the definition: suppose $\Ba'=N\Ba N^*$, then $\phi'_\alpha=N\phi_\alpha$ and therefore $M=N$.
	
	(2) This is a direct consequence of $\det(M'M)=\det(M')\det(M)$.
	
	(3) It suffices to consider the case $\Ba=\Ba'$ with two arbitrary choices of bases, and to show that $s(\Ba,\Ba')=1$. Indeed, the chain rule (2) then implies the independence of $s(\Ba,\Ba')$ on the choice of basis for $\Ba'$ by setting $\Ba''=\Ba'$, and similarly for the independence on the basis of $\Ba$.
	
	Suppose $\Ba=\Ba'$. Then $M$ is the unitary matrix describing the basis change from $\phi_\alpha$ to another basis $\phi_\alpha'$, and is of the block matrix form
	\begin{equation}\label{Mfifi}
	M=\begin{pmatrix}V&0\\0&\overline V\end{pmatrix},
	\end{equation}
	where $V_{\alpha\beta}=\braket{\phi_\alpha}{\phi'_\beta}$ for $\alpha,\beta=1,\ldots,d$, and $\overline{V}$ denotes the elementwise complex conjugate. Clearly, in this case $\det M=1$.
	
	(4) To any basis $\phi_\alpha$ for $\Ba\HH$ we can associate an $\ph$-real basis $\psi_\alpha$ for $\HH$ by setting $\psi_\alpha=(\phi_\alpha+\eta\phi_\alpha)/\sqrt2$ and $\psi_{\alpha+d}=i(\phi_\alpha-\eta\phi_\alpha)/\sqrt2$ for $\alpha=1,\ldots,d$ and the corresponding $\psi'_\alpha$ and $\psi'_{\alpha+d}$ in terms of $\phi'_\alpha$. The basis change between $\psi$ and $\phi$ and between $\psi'$ and $\phi'$ is then given by a fixed matrix $D=D'$. We then get $M=D^*\widetilde MD$, where $\widetilde M_{\alpha\beta}=\braket{\psi_\alpha}{\psi'_\beta}$ is real. Hence $\det M=\abs{\det D}^2\det\widetilde M$ is also real.
	
	(5) This is obvious, because the basis $\phi_\alpha$ can be chosen to depend continuously on $\Ba$, for example by projecting and re-orthogonalizing.
	
	(6) The if part is clear from (4) and (5). For the converse, consider the real bases $\psi$ and $\psi'$, related by $\psi'_\alpha=\sum_\beta\widetilde M_{\beta\alpha}\psi_\beta$, where $\det\widetilde M=1$. Since ${\rm SO}(2d)$ is connected, we can find a continuous family $M(t)$ with $M(0)=\idty$ and $M(1)=\widetilde M$. In this way we get a continuous family of bases $\psi(t)$. By applying $D$ we get $\phi_\alpha(t)$ and hence $\Ba(t)$ connecting $\Ba=\Ba(0)$ and $\Ba'=\Ba(1)$.
\end{proof}

This can also be written in terms of Pfaffians: Fix a real basis, and express the projection $\Ba\in\etflipp$ in terms of $U_\Ba=i\Ba-i(\idty-\Ba)$. Then, since $\ph U_\Ba\ph=U_\Ba$, $U_\Ba$ is real. Moreover, $U_\Ba^*=-U_\Ba$, so $U_\Ba$ is antisymmetric. Hence the Pfaffian $\pfaff(U_\Ba)$ of $U_\Ba$ is well defined and real. Since $\pfaff(A)^2=\det(A)$, $\pfaff(U_\Ba)$ has modulus $1$, and $\pfaff(U_\Ba)=\pm1$ for every $\ph$-flipped projection $\Ba$. That sign by itself has no meaning, because it depends on the real basis chosen. This dependence is governed by the identity $\pfaff(RAR^T)=\det(R)\pfaff(A)$, in particular for orthogonal $R$. Comparing with the proof it is clear that with an orthogonal transformation of determinant $(-1)$  one switches between the two connected components. To summarize: in the setting of the Lemma,
\begin{equation}\label{pfaffian}
\pfaff(U_\Ba)=s(\Ba,\Ba')\pfaff(U_{\Ba'}).
\end{equation}

\begin{prop}\label{prop:symDti}
	Let $W_1$ and $W_2$ be translation invariant walks with continuous bands on the same cell structure satisfying the assumptions of symmetry type $\symD$ for the same symmetry operator $\ph$. Let $\Ba_i(k)$ ($i=1,2$) denote the eigenprojections of $\Wh_i(k)$ for the upper half plane. Then $\Ba_i(0)$ and $\Ba_i(\pi)$ are $\ph$-flipped projections. Between these, consider the four signs $s(\cdot,\cdot)$ in the following diagram:
	$$
\begin{tikzpicture}
  \node (a) {$\Ba_1(0)$};
  \node (b) [right=2cm of a] {$\Ba_1(\pi)$};
  \node (c) [below=of a] {$\Ba_2(0)$};
  \node (d) [right=2cm of c] {$\Ba_2(\pi)$};
  \draw[<->] (a) -- node[above] {$s_1$} (b);
  \draw[<->] (c) -- node[below] {$s_2$} (d);
  \draw[<->] (a) -- node[left] {$s_0$} (c);
  \draw[<->] (b) -- node[right] {$s_\pi$} (d);
\end{tikzpicture}$$
	Then $W_1$ and $W_2$ are homotopic in the set of such walks if and only if $s_0=1$ and $s_\pi=1$.
	Moreover, $s_i$ determines the invariants $\sixR(W_i)$ according to the formula
	\begin{equation}\label{preberry}
	(-1)^{\textstyle\sixR(W)}=s(\Ba(0),\Ba(\pi))=\frac{\pf(W(0))}{\pf(W(\pi))}.
	\end{equation}
\end{prop}
Note that a similar formula was found by Kitaev for the Majorana number of a translation invariant gapped Hamiltonian on a finite chain \cite{KitaevMajorana}.

\begin{proof}
	(1) Clearly the condition $s_0=1=s_\pi$ is necessary for homotopy by \Qref{lem:etaflip}\,(6), since a homotopy between $W_1$ and $W_2$ necessarily needs to continuously connect $\Wh_1(k)$ and $\Wh_2(k)$, and the band projections are continuous functions of $\Wh_i(\{0,\pi\})$ into $\etflipp$.
	
	(2) For the converse we first show that, when $s_\pi=1$, i.e., when $\Ba_1(\pi)$ and  $\Ba_2(\pi)$ are homotopic in $\etflipp$, we can deform $W_1$ in such a way that we even have $\Ba_1(\pi)=\Ba_2(\pi)$, but leave $\Ba_1(0)$ unchanged.
	To this end consider a homotopy of $\Ba_1(\pi)$ and $\Ba_2(\pi)$, in the form of a continuous curve $k\mapsto V_k\in SO(2d)$, for $k\in[0,\pi]$ such that $V_0=\idty$ and $V_\pi \Ba_1(\pi)V_\pi^*=\Ba_2(\pi)$. Moreover, we extend this to negative values by setting $V_{-k}=V_k$. Then consider the walks $W_t$, $t\in[0,1]$ with
	\begin{equation}\label{whtk}
	\Wh_t(k)=V_{(1-t)k}\Wh_1(k)V_{(1-t)k}^* .
	\end{equation}
	First of all this notation makes consistent use of $W_1$ since this equation is an identity for $t=1$. Moreover, it gives a continuous family of \symD-symmetric walks,  $\Wh_0(0)=\Wh_1(0)$, and  $\Wh_0(\pi)=\Wh_2(\pi)$. The band projections are then connected as claimed. By replacing $W_1$ with $W_0$ we may hence assume without loss that
	$\Ba_1(\pi)=\Ba_2(\pi)$. With a completely analogous construction we can achieve $\Ba_1(0)=\Ba_2(0)$.
	
	(3) What we are now left with are two norm-continuous curves of rank $d$  projections in $\Cx^{2d}$ with fixed end points $\Ba_1(0)=\Ba_2(0)$ and $\Ba_1(\pi)=\Ba_2(\pi)$, where $2d$ is the local cell dimension. These can be considered as two curves on the so called Grassmannian manifold $\Grass_{d}(\Cx^{2d})$, that is, the manifold of $d$-dimensional subspaces of $\Cx^{2d}$. By \Qref{lem:Grassmann} below, this manifold is simply connected and hence the two paths of projections and therefore the walks $W_1$ and $W_2$ are homotopic.
		
	(4) For the first equality in \eqref{preberry} first note that $s_1s_\pi s_2s_0=1$, whence $s_0=s_\pi$ is equivalent to $s_1=s_2$. Now assume $s_0=1$. Then $W_1$ and $W_2$ are homotopic iff $s_\pi=1$ iff $s_1=s_2$. For the case of $s_0=-1$ let $N$ be the unitary that, in the basis of \Qref{lem:etaflip} for $\Ba=\Ba_2(0)$, swaps $\phi_1$ and $\phi_{d+1}$ and acts as the identity on the complement. Then $\ph N=N\ph$ and $\det N=-1$. Now instead of $\Wh_2$ consider $\Wh_2'=N^*\Wh_2N$.  Conjugation with $N$ leaves $s_2$ invariant, but changes both signs $s_0=s_\pi$. Hence we are left with the case above: $s_1=s_2\Leftrightarrow\sixR(W_1)=\sixR(W_2')$.
	Since $\sixR$ is also invariant under conjugation with local unitaries, $\sixR(W_2)=\sixR(W_2')$ and hence $\sixR(W_i)$ and $s_i$ label the same classes. Equation \eqref{preberry} just translates between $\Ir_2$ considered as an additive or multiplicative group, respectively. For the correct assignment of invariants consider a walk with constant $\Ba(k)=\Ba(0)$. This clearly has $s(\Ba(0),\Ba(\pi))=1$ and the corresponding walk $W$ acts locally in each cell, which, by the assumption of balanced cells, implies $\six(W)=0$.
	The second identity follows from \eqref{pfaffian}.	
\end{proof}

The following lemma is a well known result from the theory of homogeneous spaces. To give a self contained description, however, we state it here and also give a sketch of a proof, with references to more detailed descriptions.
\begin{lem}\label{lem:Grassmann}
	The Grassmannian manifold $\Grass_n(\Cx^d)$ is simply connected.
\end{lem}

\begin{proof}[Sketch of proof ]
	$\Grass_n(\Cx^d)$ is known to be isomorphic to the homogeneous space\linebreak $SU(d)/ \left(SU(n)\times SU(d-n)\right)$ \cite{Arvanitoyeorgos}. The fundamental group of this space can be computed using the exact homotopy sequence of the fibration $p\colon SU(d)\to SU(d)/(SU(n)\times SU(d-n))$, with fibre $SU(n)\times SU(d-n)$, where $p$ is the natural quotient map (for further reading see e.\,g. \cite{Steenrod}). For any such fibration
	there is the so called ``homotopy sequence of a fibration'', which is exact \cite[Thm 11.48]{Rotman}. In our case, by $\pi_0(SU(d))=\{0\}=\pi_1(SU(d))$ (considered as a single-element set or the trivial group), this gives rise to the exact sequence
	\begin{align}
		\{0\}\to\pi_1\bigl (SU(d)/(SU(n)\times SU(d-n))\bigr )\to\pi_0\bigl (SU(n)\times SU(d-n)\bigr )\to\{0\},
	\end{align}
	which implies
	\begin{align}
		\pi_1\bigl (\Grass_n(\Cx^{d})\bigr )=\pi_0\left( SU(n)\times SU(d-n)\right )=\{0\},
	\end{align}
	since $SU(n)\times SU(d-n)$ is connected.
\end{proof}

\subsubsection{Berry phase}

For continuously differentiable walks, the symmetry index is connected to the Berry phase of the upper bands \cite{Kita}:

\begin{cor}
	When $\Wh$ is continuously differentiable, $\sixR(W)$ can also be written as twice the Berry phase for the upper bands, i.e.
	\begin{equation}\label{Berryintegral}
	\sixR(W)\equiv\frac1{\pi i}\int_{-\pi}^\pi\mskip-10 mu dk\ \sum_{\alpha=1}^d \Bigl\langle\phi_\alpha(k),\,\frac{d\phi_\alpha(k)}{dk}\Bigr\rangle \quad\mod2.
	\end{equation}
\end{cor}

\begin{proof}
	The continuous differentiability of $\Wh(k)$ transfers to $\Ba(k)$. Hence, we find a continuously differentiable basis $\{\phi_\alpha(k)\}_{\alpha=1}^d$ for $\Ba(k)\HH$. By \eqref{bandsymPH} and similar to \Qref{lem:etaflip} this can be extended to a continuously differentiable basis  of $\HH$ by setting $\phi_{\alpha+d}(k)=\ph\phi_\alpha(-k)$ and therefore gives rise to a continuously differentiable family of unitaries $M(k)$, such that $\Ba(k)=M(k)\Ba M(k)^*$. Now, by \eqref{preberry}, we have $\sixR(W)=\log\det M(\pi)/(i\pi)$, where it does not matter which branch we choose, since it will be evaluated $\mod 2$. Using $\frac{d}{dk}(\log\det M(k))=\tr(M^*(k)\frac{dM(k)}{dk})$ this can be expressed as
\begin{equation}
		\sixR(W)\equiv\frac1{\pi i}\int_0^\pi\mskip-10 mu dk\sum_{n=1}^{2d}\Bigl\langle\phi_n(k),\,\frac{d\phi_n(k)}{dk}\Bigr\rangle \quad\mod2,
	\end{equation}
	which evaluates to the claimed formula, if we plug in the definition of $\phi_{\alpha+d}(k)$.
\end{proof}

\subsubsection{Completeness}

The completeness result in \cite{LongVersion} tells us that walks that have the same indices are homotopic. \Qref{prop:symDti}, however, seems to contradict this by describing two walks $W_1$, $W_2$ with the same indices as non-homotopic if $s_0=s_\pi=-1$. This difference is a consequence of the restriction to the cell structure which is fixed by the (minimal) translations that translation invariance refers to. Our general theory and the completeness result therein is entirely independent of translation invariance. Therefore, changing the assumptions by adding translation invariance with respect to a fixed cell structure one can expect additional invariants which reveal an even finer structure of the set of admissible walks. As \Qref{prop:symDti} shows, this is indeed the case. Yet, if we only demand homotopies to respect translation invariance with respect to shifts by an even number of sites, e.g. by regrouping neighbouring even and odd sites as if putting up domino tiles (see \Qref{fig:regroupeddomino} and \Qref{sec:regrouping}), the additional invariant is rendered useless, i.e. the regrouped walks become homotopic on the coarser lattice:

\begin{lem}\label{lem:dominoti}
	Let $W_1$ and $W_2$ be translation invariant walks with continuous bands on the same cell structure satisfying the assumptions of symmetry type $\symD$ for the same symmetry operator $\ph$.
Assume that $\sixR(W_1)=\sixR(W_2)$, which by \Qref{prop:symDti} means that $s_1=s_2$. This does not fix the values of $s_0$ and $s_\pi$, but only their equality. 	Then:
\begin{itemize}
\item[(1)] After regrouping neighbouring cells pairwise, the regrouped walks $W_{1,r}$ and $W_{2,r}$ are homotopic.
\item[(2)] There are trivial (i.e., cell-wise acting), \symD-admissible walks $W_1^0$ and $W_2^0$
     such that $W_1\oplus W_1^0$ and $W_2\oplus W_2^0$ are homotopic.
\end{itemize}
\end{lem}

\begin{proof}
(1) Let $s=s_0=s_\pi$, which are equal by $s_1s_2 s_0s_\pi=1$ and $s_1=s_2$.
If $s=1$, $W_1$ and $W_2$ are homotopic on the original lattice by \Qref{prop:symDti}. This transfers to the regrouped lattice in the obvious way.
Suppose now, that $s=-1$. As a first step transform $W_1$ and $W_2$ into their corresponding flat-band, as described in \Qref{lem:TIflat}. We get $\Wh_j(k)=i(2\Ba_j(k)-\idty)$, with $j=1,2$. Then, by \Qref{lem:etaflip}\,(1), there are unitaries $N,M$, commuting with $\ph$, with $\det N=\det M=-1$ s.t. $\Wh_2(0)=N\Wh_1(0)N^*$ and $\Wh_2(\pi)=M\Wh_1(\pi)M^*$.
Relabelling a walk $W$ as shown in \Qref{fig:regroupeddomino} leads to a regrouped walk $W_r$ which has Fourier transform $\Wh_r(k)$ as given by \eqref{Whregroup}.
Then, if $W_1$ and $W_2$ are related as above, $\Wh_{1,r}(0)$ and $\Wh_{2,r}(0)$ are related by conjugation with $N_r=H(0)(N\oplus M)H(0)^*$. Hence $s_{0,r}=\det(N_r)=\det N\det M=1$.
For $s_{\pi,r}$, note that by \eqref{preberry} $s_1$ and $s_2$ are invariant under regrouping, since the right symmetry index $\sixR$ does not change under such operation. Therefore, by $s_0s_1s_\pi s_2=1$, which also holds for the regrouped walks, we get $s_{\pi,r}=s_{0,r}=1$ and hence $W_1$ and $W_2$ are connected by a period-2 homotopy.

(2) Now let $W_i^0=\bigoplus_\Ir \Wh_i(0)$ be the walk, which is block diagonal, with $\Wh_i(0)$ acting locally in each cell and $\widetilde W_i=W_i\oplus W_i^0$. Then $W_i^0$ clearly fulfils the right symmetry condition and has trivial symmetry index ($s_i=1$).
We get $\widetilde s_0=s_0^2=1$ ($=\widetilde s_\pi$) and hence $\widetilde W_1$ and $\widetilde W_2$ are homotopic by \Qref{prop:symDti}.
\end{proof}

\subsubsection{Example: Building bridges}

The phenomenon of period two homotopies for walks of type $\symD{}$ also occurs in the split-step walk $W(\theta_1,\theta_2)$ (see \Qref{sec:splitstep}). As the phase diagram indicates by the white lines between the plaquettes, it is not possible to continuously connect two split-step walks from two plaquettes with the same phase, e.\ g.\ $W_0=W(\theta_1,\theta_2)$ and $W_1=W(\theta_1,\theta_2+\pi)$, by varying the two parameters $\theta_1,\theta_2$. Also leaving the class of split-step walks while keeping the cell structure fixed does help: to see this, we restrict ourselves to flat-band walks (centers of the plaquettes), since any two walks can always be deformed to the respective flat-band candidates in their plaquettes. We then get $W_0=-W_1$, which for flat-band walks implies
\begin{equation}\label{eq:bridgeprojections}
\Ba_1(k)=\idty-\Ba_0(k).
\end{equation}
The intertwining unitary for $k=0,\pi$ is then always of the form $M=U\sigma_xU^*$, which yields $s_0=s_\pi=-1$. Hence, sticking to two dimensional cells, $W_0$ and $W_1$ are not homotopic. If we now regroup the cell structure, \eqref{eq:bridgeprojections} remains true. But now an intertwining unitary is of the form $M=U(\sigma_x\otimes\idty_2)U^*$, which has $\det M=+1$.

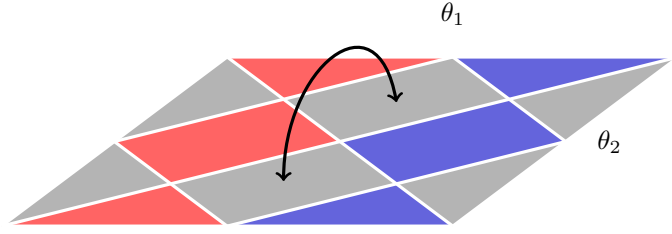
\begin{figure}[t]
	\begin{center}
%
%
  \begin{tikzpicture}[scale=1.5,=>stealth]
  
    \definecolor{pmcol}{RGB}{180,180,180}
	\definecolor{nncol}{RGB}{180,180,180}
	\definecolor{nmcol}{RGB}{100,100,220}
	\definecolor{pncol}{RGB}{255,100,100}
    
    \fill[pncol] (-3,-1) -- (-1.5,-.629) -- (-1,-1);	
	\fill[pncol] (-1.5,-.629) -- (0,-.25) -- (-.5,.125) -- (-2,-.25);
    \fill[pncol] (-.5,.125) -- (-1,.5) -- (1,.5);
    
    \fill[nmcol] (-1,-1) -- (.5,-.629) -- (1,-1);	
	\fill[nmcol] (.5,-.629) -- (0,-.25) -- (1.5,.125) -- (2,-.25);
    \fill[nmcol] (1.5,.125) -- (1,.5) -- (3,.5);
    
    \fill[nncol] (-3,-1) -- (-1.5,-.629) -- (-2.,-.25);	
	\fill[nncol] (-2,-.25) -- (-1,.5) -- (-.5,.125);
    
    \fill[nncol] (-1,-1) -- (-1.5,-.629) -- (0,-.25) -- (.5,-.629);
    \fill[nncol] (0,-.25) -- (-.5,.125) -- (1,.5) -- (1.5,.125);
    
    \fill[nncol] (1,-1) -- (.5,-.629) -- (2,-.25);
    \fill[nncol] (2,-.25) -- (1.5,.125) -- (3,.5);
    

    \draw[white,very thick] (-3,-1) -- (3,.5);
    \draw[white,very thick] (-1,.5) -- (1,-1);
  
    \draw[white,very thick] (-3,-1) -- (-1,.5);
    \draw[white,very thick] (3,.5) -- (1,-1);

    \draw[white,very thick] (-1,.5) -- (3,.5);
    \draw[white,very thick] (-3,-1) -- (1,-1);

    \draw[white,very thick] (-2,-.25) -- (1,.5);
    \draw[white,very thick] (-1,-1) -- (2,-.25);

    \draw[white,very thick] (-2,-.25) -- (-1,-1);
    \draw[white,very thick] (2,-.25) -- (1,.5);

    \draw[<->,very thick] (-.5,-.6) .. controls +(90:1) and +(100:1) .. (.5,.1);
    
	\draw (2.2,-.25)  node[right,align=center]{{$\theta_2$}};   
	\draw (1,0.7) node[above,align=center]{ {$\theta_1$}};

  \end{tikzpicture}
\end{center}
	\caption{The ``Bridge over troubled gap closings'' vizualizing the homotopy \eqref{eq:bridge} of trivial walks with respect to the period-two regrouping. Without the regrouping, and hence the breaking of translation symmetry, all symmetry preserving paths would have to contain a gapless walk, because the end points of the bridge are related by $s_0=-1$. In particular, paths in the split-step parameter plane (\Qref{fig:harlequin}) have to cross a white line.}
\end{figure}

For an explicit example consider $W_0=W(0,\pi/2)=-i\sigma_y$ and $W_1=W(0,-\pi/2)=i\sigma_y$. A homotopy between the two regrouped versions is then given by
\begin{equation}\label{eq:bridge}
W_t=%
\begin{pmatrix}
0 & -\cos(\pi t) & -\sin(\pi t) & 0\\
\cos(\pi t) & 0 & 0 & \sin(\pi t)\\
\sin(\pi t) & 0 & 0 & -\cos(\pi t)\\
0 & -\sin(\pi t) & \cos(\pi t) & 0
\end{pmatrix}.
\end{equation}

\subsection{Chiral reduction}\label{sec:chiralFred}

In contrast to $\symD{}$, the symmetry types $\symAIII{}$, $\symBDI{}$, $\symCII{}$ and $\symDIII{}$ all contain a chiral symmetry.  The admissibility condition for this chiral symmetry leads to a particular structure for admissible operators with respect to the eigenbasis of $\ch$. Combining this structure with the flattening of the bands (see \Qref{lem:TIflat} and the accompanying discussion) leads to a process which we call the \emph{chiral reduction}. We begin with discussing the structure of chirally symmetric unitaries:
\begin{lem}\label{lem:chiralblocks}
	Let $\HH$ be a Hilbert space
	equipped with a chiral symmetry $\ch$ with $\ch^2=(-1)^s\idty,s\in\{0,1\}$, whose $\pm i^s$-eigenspaces $\HH_\pm$ have the same dimension.
	Let $U$ be a unitary operator which is admissible for this symmetry, i.\,e.\ $\ch U=U^*\ch$.
	With respect to the decomposition $\HH=\HH_+\oplus\HH_-$, these operators can be written in blocks as
	\begin{align}\label{eq:blocks}
		\ch=i^s\begin{pmatrix}\idty&0\\0&-\idty\end{pmatrix}  \ \mbox{and}\
		U=\begin{pmatrix} A&B\\C&D\end{pmatrix}
	\end{align}
	with
	\begin{align}\label{chiralstructure}
		A&=A^*  &  BB^*+A^2&=\idty  \nonumber\\
	 	D&=D^*  &  B^*B+D^2&=\idty\\
	 	C&=-B^* &  AB-BD&=0 \nonumber
	\end{align}
	Then
	\begin{itemize}
		\item[(1)]  $B$ has a bounded inverse iff $U$ has proper gaps at $1$ and $-1$.
		\item[(2)] $B$ is Fredholm iff $U$ has an essential gap.
		\item[(3)] $U$ is a flat-band unitary, i.\,e.\ $U^2=-\idty$, iff $A=D=0$ and $B$ is unitary.
	\end{itemize}
\end{lem}

Note that by the assumption that each cell is balanced, the condition that the dimensions of $\HH_+$ and $\HH_-$ agree is automatically fulfilled in our walks setting.

\begin{proof}
	The algebraic relations between the blocks are direct consequences of the admissibility and the unitarity.
	Consider the hermitian operator
	\begin{equation}\label{fredIm}
	\im U=\frac1{2i}(U-U^*)=\begin{pmatrix}0&-iB\\iB^*&0\end{pmatrix}.
	\end{equation}
	Clearly, it is invertible iff $B$ is invertible. Moreover, this is equivalent to the absence of any eigenvalues $z$ or other spectrum in a strip $\abs{\im z}<\veps$. This proves (1). Stating (1) for the image $\calk U$ of $U$ in the Calkin algebra is exactly (2). (3) is a direct consequence of the flat-band condition $U=-U^*$.
\end{proof}

For the homotopy classification it turns out to be useful to consider flat-band walks only, since the classification of admissible unitaries $U$ then reduces to the classification of the upper right matrix blocks $B$, which are also unitary and have an effectively smaller symmetry group.
Note that the flat-band condition $U^2=-\idty$ guarantees a proper gap for $U$. Therefore we do not need to impose such condition on $B$. Since the flattening procedure will be important when we discuss completeness of the invariants in \Qref{sec:chiralcomplete}, we will take a closer look at it in the context of the chiral eigenbasis in \Qref{sec:chiralflat}.

In case of $\symAIII{}$ no further symmetries are present and the problem reduces to the homotopy classification of essentially local unitaries $B$ with no further restriction. As will be made explicit later, this task is completely covered by the index for quantum walks (see \Qref{sec:bands}, \cite{OldIndex}).
Adding a second (and hence also a third) symmetry yields different conditions on $B$, depending on the symmetry type this new symmetry belongs to. We will always choose to add the particle-hole symmetry $\ph$. For the cases $\symBDI{}$, $\symDIII{}$ and $\symCII{}$ $\ph$ takes the form
\begin{equation}\label{eq:addph}
	\ph=\begin{pmatrix}
	\ph' & 0\\0 & \ph'
	\end{pmatrix}\qquad\text{or}\qquad
	\ph=\begin{pmatrix}
	0 & \ph'\\\ph' & 0
	\end{pmatrix},
\end{equation}
where the first case is present for the symmetry types $\symBDI{}$ (with $\ph'^2=\idty$) and $\symCII{}$ (with $\ph'^2=-\idty$) and the second case covers type $\symDIII{}$ (with $\ph'^2=\idty$). The new symmetry conditions for $B$ are then $\ph'B\ph'^*=B$ for $\symBDI{}$ and $\symCII{}$ and $\ph'B\ph'^*=-B^*$ for $\symDIII{}$. The block form of $\ph$ is a consequence of the commutation relations between the symmetries: In order to commute with the chiral symmetry in its eigenbasis, $\ph$ has to either leave the $\ch$-eigenspaces invariant (for $\ch^2=\idty$) or swap them (for $\ch^2=-\idty)$. Of course, a priori the two blocks of $\ph$ might not be the same, but we can always choose bases in the $\ch$-eigenspaces separately to obtain \eqref{eq:addph}. The three reducible cases are collected in Tab.~\ref{tab:reduction}.

The chiral reduction seems to raise a contradiction in the cases of $\symBDI{}$ and $\symCII{}$, since they seem to reduce to $\symD{}$ and $\symC{}$, which both have different index groups than \symBDI and \symCII. This contradiction is resolved by the gap condition. Since we don't have to impose this on $B$, more homotopies are allowed, and hence the classification changes. In the case of $\symDIII{}$ the chiral reduction leads to a new symmetry type, which is not contained in the tenfold way. This raises the question which other new types might be considerable, a question that will be tackled in \cite{Usym}.

\begin{table}[]\begin{center}
		\begin{tabular}{|c||c|c|c|}
				$\symS{}$	&	$\symBDI{}$	&	$\symDIII{}$	&	$\symCII$	 \\\hline
				$\symS{}'$ &  $\ph'B\ph'^*=B$ & $\ph'B\ph'^*=-B^*$ & $\ph'B\ph'^*=B$ \\
				$\ph'^2$	&	$\idty$		&	$\idty$			&	$-\idty$
		\end{tabular}
	\end{center}\caption{\label{tab:reduction} Reduction procedure for chiral symmetry types.}
\end{table}

\subsection{Index formulas for chiral symmetric walks}\label{sec:chiralformulas}

Before we start the discussion of index formulas for translation invariant walks, let us first sharpen a result from \cite{LongVersion}. Recall that a bounded operator $A$ is said to be a Fredholm operator, if it is invertible up to a compact error, i.e., there is $B$ with $AB-\idty$ and $BA-\idty$ both compact.
In this case the Fredholm index of $A$ is defined as
\begin{equation}\label{fredind}
\indF A=\dim\ker A-\dim\ker A^*.
\end{equation}

The following Lemma connects the symmetry index of operators of symmetry type  \symAIII{}, \symBDI{} or \symCII{} to the Fredholm index of a certain matrix block. The symmetry index of such operators $U$ is given by
\begin{equation}
	\six(U)=\tr_\mathcal N\ch,
\end{equation}
where $\mathcal N=\ker(U-U^*)$ \cite{LongVersion}.

\begin{lem}\label{lem:chiralFred} Let $U$ be an essentially local unitary operator of symmetry type $\symAIII{}$, $\symBDI{}$ or $\symCII{}$ with an essential gap at $\pm1$ (not necessarily flat-band) in the form \eqref{eq:blocks}. Then
	\begin{equation}\label{eq:fredindex}
	\six(U)=-{\indF}(B).
	\end{equation}
\end{lem}

\begin{proof}
The kernel of the matrix on the right hand side of \eqref{fredIm} is the set of vectors of the form $\phi_1\oplus\phi_2$ such that $B^*\phi_1=0$ and $B\phi_2=0$, i.e., the space
\begin{equation}\label{eq:chiralker}
\mathcal N=\ker(U-U^*)=\ker(B^*)\oplus\ker(B).
\end{equation}
The chiral operator $\ch$ acts on the first summand as $+\idty$ and on the second as $-\idty$, so that on this subspace
\begin{equation}
\six(U)=\tr_{\mathcal N}\ch=\dim\ker(B^*)-\dim\ker(B)=-\indF(B).
\end{equation}
\end{proof}

For symmetry type $\symCII{}$ this automatically yields an even number since the reduced symmetry with $\ph'^2=-\idty$ forces the kernel of $B$ (and $B^*$) to be even dimensional \cite{Wigner1}.
Note, that \eqref{eq:fredindex} is a generalized version of the index for quantum walks \cite{OldIndex,LongVersion}, which provides a complete homotopy classification for essentially local unitaries without any gap condition and without considering symmetries.

Let us now consider translation invariant walks with continuous band structure. Then, by assumption the chiral symmetry acts cell-wise. So we have two $d$-dimensional eigenspaces for each cell, in which $\ch$ acts as in \eqref{eq:blocks}. This structure survives Fourier transformation, so that
\begin{equation}\label{chiralWh}
\Wh(k)=\begin{pmatrix}\Ah(k)&\Bh(k)\\ -\Bh^*(k)&\Dh(k)\end{pmatrix}.
\end{equation}
The essential gap assumption demands that each $\Wh(k)$ has a gap, so by \Qref{lem:chiralblocks} $\Bh(k)$ is non-singular for all $k$. Continuity of $\Wh(k)$ trivially transfers to continuity of $\Bh(k)$, and therefore, the formula \eqref{windingcontinuous} in the following proposition makes sense. It connects our symmetry index $\sixR$ and our formulation of the bulk-boundary-correspondence to the earlier literature \cite{Asbo1}, where the formula was already given.
\begin{prop}\label{prop:Fredformula}
	Let $W$ be a translation invariant walk with continuous bands, satisfying the assumptions of \Qref{sec:nonti}, of symmetry type
	\symAIII, \symBDI{} or \symCII. Then
	\begin{equation}\label{windingcontinuous}
	\sixR(W)=\wind\bigl(k\mapsto\det \Bh(k)\bigr),
	\end{equation}
	where $\wind$ denotes the winding number of an origin-avoiding $2\pi$-periodic continuous function in the complex plane. When $\Wh$ is continuously differentiable this can also be written as
	\begin{equation}\label{windingintegral}
	\sixR(W)=\frac1{2\pi i}\int_{-\pi}^\pi\mskip-10 mu dk\ \tr\Bigl(\Bh^{-1}(k)\frac{d\Bh(k)}{dk}\Bigr).
	\end{equation}
\end{prop}

\begin{proof}
	Let $P=\Pg0$ be the half space projection, then $\sixR(W)=\six(PWP)$ \cite{LongVersion}. We now apply \Qref{lem:chiralFred} to the essentially unitary operator $PWP$ to get $\sixR(W)=-\indF(PBP)$. Now the half-space compression of a translation invariant operator (in this case $B$) is a Toeplitz operator, which is Fredholm iff its ``symbol'' $\Bh(k)$ is invertible. By a classic result \cite{Gohberg,Avron2001} the Fredholm index is then the winding number of its determinant relative to the origin. This is the formula given in the proposition.
	
	In the differentiable case we can represent the winding number of $f$ as the integral of the logarithmic derivative of $f$, and use the differentiation formula for determinants to get \eqref{windingintegral}.
\end{proof}

For an interactive tool, illustrating \eqref{windingcontinuous} for the split-step walk, see \cite{sse}.

\subsubsection{Flattening the band structure} \label{sec:chiralflat}

The flattening of the band structure, described in \ref{sec:bands} can also be done directly in the basis of \Qref{lem:chiralblocks}.
Since $B$ is non-singular, its polar isometry is unitary. The polar isometry also shares the symmetry conditions with $B$. Hence, continuously interpolating between $B$ and its polar isometry, together with an appropriate deformation of $A$ and $D$ does the job. This raises the question, if the flattening construction from \Qref{lem:TIflat} and the one sketched above are compatible. Indeed, they are essentially the same:
Consider a translation invariant walk in the chiral eigenbasis (see \eqref{chiralWh}). Then, it suffices to flatten the finite dimensional matrix $\Wh(k)$ for each $k$ (we will omit the $k$ dependence and the hat for readability). Let $B=\sum_{i=1}^{d}\beta_i\ketbra{\phi_i}{\psi_i}$
be the singular value decomposition of $B$, where $\{\phi_i\}$ ($\{\psi_i\}$) is an orthonormal basis for $\HH_+$ ($\HH_-$) and since $B$ is non-singular we have $\beta_i>0$. For non-degenerate $\beta_i$'s \eqref{chiralstructure} immediately implies $A$ ($D$) to be diagonal in the basis $\{\phi_i\}$ ($\{\psi_i\}$), with eigenvalues $a_i=d_i=\pm\sqrt{1-\beta_i^2}$. Since we can diagonalize $A$ ($D$) in each degenerate block this is also true for degenerate singular values $\beta_i$. To summarize: we have
\begin{equation}
	B=\sum_{i=1}^{d}\beta_i\ketbra{\phi_i}{\psi_i},\  A^2=\sum_{i=1}^{d}(1-\beta_i^2)\ketbra{\phi_i}{\phi_i}\ \text{and}\ D^2=\sum_{i=1}^{d}(1-\beta_i^2)\ketbra{\psi_i}{\psi_i}.
\end{equation}
The eigenvalues $\lambda_n$ of $W$ then evaluate to
\begin{equation}
	\lambda_n=\pm\sqrt{1-\beta_i^2}\pm i \beta_i.
\end{equation}
Hence, flattening the band structure by deforming the eigenvalues to $\pm i$ is equivalent to deforming all $\beta_i$ to $1$ and therefore to deforming $B$ to its polar isometry.

Having established the explicit flattening construction in the chiral eigenbasis, let us use the structure to connect the index for chiral symmetric walks with the index for particle-hole walks:

\begin{cor}\label{cor:chiralberry}
	When $\Wh$ is a continuously differentiable flat-band walk $\sixR(W)$ can also be written as twice the Berry phase for the upper band, i.e.\
	\begin{equation}
	\sixR(W)=\frac2{2\pi i}\int_{-\pi}^\pi\mskip-10 mu dk\ \sum_{\alpha=1}^d \Bigl\langle\phi_\alpha(k),\,\frac{d\phi_\alpha(k)}{dk}\Bigr\rangle.
	\end{equation}
\end{cor}

\begin{proof}
	For flat-band walks, the eigenvectors corresponding to eigenvalues $+i$ can be chosen to be of the form
	\begin{equation}
	\phi_\alpha(k)=1/\sqrt{2}\begin{pmatrix}
	-i\Bh(k)\chi_\alpha\\
	\chi_\alpha
	\end{pmatrix},
	\end{equation}
	where $\{\chi_\alpha\}_{\alpha=1}^d$ is any orthonormal basis for $\HH_-$ (independent of $k$). Now in \eqref{windingintegral} $\tr\Bigl(\Bh^*(k)\frac{d\Bh(k)}{dk}\Bigr)$ evaluates to
	\begin{equation}
		\tr\Bigl(\Bh^*(k)\frac{d\Bh(k)}{dk}\Bigr)=\sum_{\alpha=1}^d\Bigl\langle\Bh(k)\chi_\alpha,\frac{d\Bh(k)}{dk}\chi_\alpha\Bigr\rangle=2\sum_{\alpha=1}^d\Bigl\langle\phi_\alpha(k),\frac{d\phi_\alpha(k)}{dk}\Bigr\rangle,
	\end{equation}
	which is twice the Berry connection for the upper band.
\end{proof}

\subsubsection{Completeness}\label{sec:chiralcomplete}

For symmetry type $\symAIII{}$ the index classification of translation invariant walks with a fixed cell structure is complete. Each walk $W$ defines a loop $\Wh(k)$ in the unitary group $U(2d)$, which in the flat-band case is completely characterized by the unitary loop $\Bh(k)\in U(d)$. There are then no further restrictions on $\Bh(k)$ and hence completeness follows from the classic result, that two unitary loops are homotopic iff the winding numbers of their determinants coincide. Such continuous deformation automatically preserve locality on the way.
In the case of strict locality this was already shown in \cite{OldIndex}. For essentially local walks this follows from \Qref{pro:esslocalbands}, as the continuity of a loop guarantees for the essential locality of the corresponding walk.

For the symmetry types $\symBDI{}$ and $\symCII{}$ we also have to take care of the particle-hole symmetry. Let $c\colon[-\pi,\pi)\to\Cx,\,c(k)=\det(\Bh(k))$. The symmetry condition on $B$ (see Tab.~\ref{tab:reduction}) then implies
\begin{equation}\label{eq:phcondition}
	c(k)=\overline{c(-k)}.
\end{equation}

The effect of this restriction is, that similar to $\symD{}$, $\sixR$ is no longer complete
for a fixed cell structure. Indeed, consider e.\,g.\ the $\symBDI{}$-symmetric walks with constant loops $B_1(k)=1$ and $B_2(k)=-1$, which both yield walks with trivial symmetry index on the cells $\HH_x=\Cx^2$. But it is clearly not possible to transform them into each other, without violating the gap condition for the corresponding walks (\Qref{lem:chiralblocks}), since such path would have to cross the origin for $k\in\{0,\pi\}$. This example indicates an additional invariant, similar to the \emph{regrouping-invariant} in case $\symD{}$.

\begin{lem}\label{lem:BDIcomplete}
	Let $W_1$ and $W_2$ be two translation invariant walks with continuous bands on the same cell structure. Assume $W_1$ and $W_2$ to be admissible for the same symmetries of type $\symBDI$, with $\wind(c_1)=\wind(c_2)$, where $c_i(k)=\det\Bh_i(k)$. Then $W_1$ and $W_2$ are homotopic iff $\sign(c_1(0))=\sign(c_2(0))$.
\end{lem}

\begin{proof}
As a first step, we deform both walks to their respective flat-band forms to reduce the problem to the upper right blocks $\Bh_i(k)$ in the chiral eigenbasis. Since $\Bh_i(k)$ is then unitary, $c_i(k)$ lies on the unit circle for all $k$. In particular we have $c_i(k)=\exp(2\pi ia_i(k))$, with $a(k)\in\Rl$ and $a_i(\pi)=a_i(-\pi)+\wind(c_i)$. Moreover the symmetry condition \eqref{eq:phcondition} implies $c_i(k)=\pm1$ for $k\in\{0,\pm\pi\}$ and the corresponding $\Bh_i(k)$ to be orthogonal with respect to an $\ph'$ invariant basis. In order to deform $\Wh_1$ into $\Wh_2$ it is now enough to only consider the paths $k\mapsto c_i(k)$ for $k\in[0,\pi]$, since the second half of the loop is then determined by \eqref{eq:phcondition} and we have to show, that these two paths are homotopic. To do so we need that the two endpoints can be connected by a continuous path inside the set of orthogonal matrices. A necessary and sufficient condition for this is $c_1(0)=c_2(0)$ and $c_1(\pi)=c_2(\pi)$, since the connected components of the orthogonal group are labeled by the values $\pm1$ of the determinant. If this is fulfilled, we are left with a closed loop in the unitary group, which by $\wind(c_1)=\wind(c_2)$ does not wind around the origin and is therefore contractible. Moreover, if $c_1(k=0)=c_2(k=0)$, the same already follows for $k=\pi$: by symmetry it is $a(k)+a(-k)=\text{const.}=2a(0)$ and therefore $\wind(c)=2(a(\pi)-a(0))$. This gives $c(\pi)=(-1)^pc(0)$, with $p=\wind(c)\mod 2$, which is the same for both $c_i$.
\end{proof}

Similar to the case of symmetry type \symD, the additional invariant can be trivialized, by either regrouping neighbouring cells once or by adding trivial systems to the respective walks under consideration:

\begin{lem}\label{lem:BDIregroup}
	Consider the setting of \Qref{lem:BDIcomplete}.	Then:
	\begin{itemize}
		\item[(1)] After regrouping neighbouring cells pairwise, the regrouped walks $W_{1,r}$ and $W_{2,r}$ are homotopic.
		\item[(2)] There are trivial (i.e., cell-wise acting), \symBDI-admissible walks $W_1^0$ and $W_2^0$
		such that $W_1\oplus W_1^0$ and $W_2\oplus W_2^0$ are homotopic.
	\end{itemize}
\end{lem}

\begin{proof}
	If we regroup a given walk once, according to \eqref{Whregroup}, we get $c_r(k)=c(k/2)c(k/2+\pi)$ and hence $c_r(0)=c(0)c(\pi)=c(0)^2(-1)^p=(-1)^p$, which only depends on the winding number and is therefore the same for $W_1$ and $W_2$.
		
	Now let $W_i^0=\bigoplus_\Ir \Wh_i(0)$ be the walk, which is block diagonal, with $\Wh_i(0)$ acting locally in each cell. It fulfils the right symmetry and has trivial symmetry index. We get $\widetilde c_i(0)=c_i(0)^2>0$, where $\widetilde c_i(k)$ denotes the determinant of the upper right chiral block of $W_i\oplus W_i^0$. Now, since $\sign(\widetilde c_1(0))=\sign(\widetilde c_2(0))$, $ W_1\oplus W_1^0$ and $W_2\oplus W_2^0$ are homotopic.
\end{proof}
Note that the construction of the trivial walks is exactly the same as for symmetry type \symD (\Qref{lem:dominoti}).

For symmetry type $\symCII{}$ the additional regrouping invariant does not appear:
\begin{lem}\label{lem:CIIcomplete}
	Let $W$ be a translation invariant walk with continuous bands of symmetry type $\symCII{}$. Then $\sign(c(0))=\sign(c(\pi))=1$. Moreover, two translation invariant walks $W_1,W_2$, which are admissible for the same symmetries of type $\symCII{}$ on the same cell structure are homotopic iff $\wind(c_1)=\wind(c_2)$.
\end{lem}
\begin{proof}
	Consider again the flat-band case, where $\Bh(k)$ is unitary. For $k\in\{0,\pi\}$, $\Bh(k)$ commutes with $\ph'$, with $\ph'^2=-\idty$. This implies $\det(\Bh(0))=\det(\Bh(\pi))=1$: Let $B\phi=\lambda\phi$, with $B$ finite dimensional and commuting with $\ph'$, Then $B\ph'\phi=\overline\lambda\ph'\phi$ such that the non-real eigenvalues of $B$ come in complex conjugate pairs. Since $\ph'^2=-\idty$ any real eigenvalue occurs with an even multiplicity and we get $\det(B)=1$.
	
	The rest of the proof is now similar to the proof of \Qref{lem:BDIcomplete} and the only thing we need to show is, that the set of finite dimensional unitaries which commute with a given $\ph'$ is connected. Consider such a unitary $B$. Since the eigenvalues of $B$ come in complex conjugate pairs, we can continuously shift each pair to $+1$, without changing the eigenvectors, and thus keeping the symmetry. Hence, every such unitary is connected to the identity. Note that the only difference to the reduced symmetry condition for symmetry type $\symBDI{}$ is the even dimensional $-1$-eigenspace which makes such deformations possible.
\end{proof}

\subsection{Symmetry type \symDIII{}}
In the case of symmetry type $\symDIII{}$, \Qref{lem:chiralFred} does not apply. In fact, due to the effective symmetry of $B$ after the chiral reduction process ($\ph'B\ph'^*=-B^*$), the Fredholm index of $B$ always evaluates to zero. For such operators, there is, however, a similar invariant \cite{SchulzZ2}. It is defined for odd symmetric Fredholm operators, i.\,e.\ Fredholm operators $T$ satisfying $(IK)T(IK)^*=T^*$, with $I$ being a real unitary with $I^2=-\idty$, and $K$ the complex conjugation with respect to a suitable basis. For such operators the index is defined as the parity of the dimension of the kernel, i.e.
\begin{equation}
	\ind_2(T)=\dim(\ker(T))\:\mod\ 2,
\end{equation}
which is a complete homotopy invariant and invariant under compact perturbations that respect the odd symmetry \cite{SchulzZ2}. The following Lemma connects the symmetry index of a unitary operator of symmetry type $\symDIII{}$ given by $\six(U)=\dim\mathcal N\ \mod\ 4$, with $\mathcal N$ defined in \eqref{eq:chiralker}, with $\ind_2$:

\begin{lem}\label{lem:DIIIfred} Let $U$ be an essentially gapped unitary of symmetry type $\symDIII{}$. Then
	\begin{equation}\label{eq:SBindex}
		\six(U)=2\ \ind_2(I^*B),
	\end{equation}
where $I$ is a real unitary with $I^2=-\idty$ and $B$ is the upper right block of the decomposition given in \eqref{eq:blocks}.
\end{lem}

\begin{proof} From the \symDIII-admissibility condition in Table \ref{tab:reduction}, \eqref{eq:chiralker} yields
	\begin{equation}
		\dim \mathcal N=2\ \dim\ker(B),
	\end{equation}
and hence $\six(U)=2\dim\ker(B)\ \mod\ 4$. By \cite[Proposition 1]{SchulzZ2}, an operator $T$ is odd symmetric with respect to $I$, iff $T=I^*\widetilde B$, for a skew-symmetric operator $\widetilde B$. Choosing a basis in which $\ph'=K$ we have $B^T=-B$, where $B^T$ denotes the matrix transpose with respect to this basis. Thus, in this basis $I^*B$ is odd symmetric with respect to any real unitary $I$ with $I^2=-\idty$. By the invertibility of $I$ we have $\dim\ker(B)=\dim\ker(I^*B)$, which completes the proof.
\end{proof}

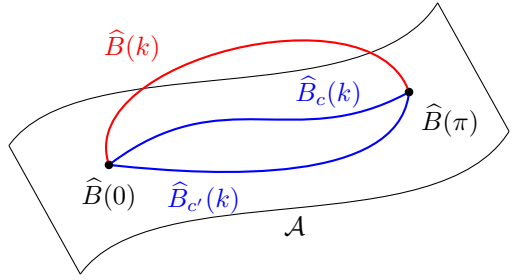
\begin{figure}[]
	\begin{center}\begin{tikzpicture}[scale=1.9]
\def\dx{-0.5}
\def\dy{0.85}
\def\dummy{.05}
\draw (0,0) .. controls (0.5,0.7) and (2.5,0.2) .. node[below]{$\mathcal A$} (3,1);
\draw (\dx,\dy+\dummy) .. controls (0.5+\dx,0.7+\dy+\dummy) and (2.5+\dx,0.2+\dy+\dummy) ..  (3+\dx,1+\dy+\dummy);
\draw (0,0) -- (\dx,\dy+\dummy);
\draw (3,1) -- (3+\dx,1+\dy+\dummy);

\node[circle,draw,label=south:$\Bh(0)$,fill,scale=0.3] (w0) at (0.2,0.9*\dy){};
\node[circle,draw,label=south east:$\Bh(\pi)$,fill,scale=0.3] (wpi) at (2.3,1.5*\dy) {};

\draw[red,thick] (w0) .. controls (0,1.5) and (2,2) .. node[pos=0.35,above left]{$\Bh(k)$} (wpi);
\draw[blue,thick] (w0) .. controls (1,1.6*\dy) and (1.5,1*\dy) .. node[near end,above]{$\Bh_c(k)$} (wpi);
\draw[blue,thick] (w0) .. node[near start,below]{$\Bh_{c'}(k)$} controls (1,0.8*\dy) and (2.2,0.75*\dy) .. (wpi);
\end{tikzpicture}\end{center}
	\caption{Visualization of two antisymmetric closures. The path $\Bh(k)$, for $k\in[0,\pi]$ is depicted in red. The two different antisymmetric closures $\Bh_c(k)$ and $\Bh_{c'}(k)$, for $k\in[\pi,2\pi]$, are depicted in blue. Their difference is a closed path inside the antisymmetric manifold (blue loop).}
	\label{fig:d3flyingcarpet}
\end{figure}

In \cite[Sect. 4]{SchulzZ2} also an odd symmetric Gohberg-Krein theorem is derived. It states, that $\ind_2$ of an odd symmetric Toeplitz operator (with continuous symbol, compare with \Qref{prop:Fredformula}) is equal to the spectral flow of its symbol modulo $2$.

Here we give an alternative derivation of the right symmetry index of \symDIII symmetric walks in terms of a winding number. In order to derive the index formula, we combine ideas from the two previous sections. Roughly speaking, we on the one hand start with the same quantity as for other chiral walks ($\det(\Bh)$) but on the other, only the path for $k\in[0,\pi]$ matters, similar to symmetry type \symD.

\begin{prop}\label{prop:D3complete}
	Let $W$ be a translation invariant walk of symmetry type $\symDIII{}$ with continuous bands, and let $\Bh(k)$ be the component of the chiral reduction \eqref{chiralWh}. Then, there is a continuous closed curve $[0,2\pi]\ni k\mapsto\Bh_c(k)$ of operators such that $\Bh(k)=\Bh_c(k)$ for $k\in[0,\pi]$ and $\Bh_c(k)$ antisymmetric and non-singular for $k\in[\pi,2\pi]$. Let $c(k)=\det\Bh_c(k)$ be the corresponding closed curve in $\Cx$.
Then $\wind(c)\ \mod\ 2$ is a complete homotopy invariant and
	\begin{equation}\label{DIIIwinding}
	\sixR(W)\equiv 2\ \wind(c)\quad\mod4.
	\end{equation}	
\end{prop}

\begin{proof}
For a translation invariant walk $\Wh(k)$ the reduced symmetry condition for $\Bh$ gives $\Bh(k)=-\Bh(-k)^T$, in an $\ph'$- invariant basis. Hence $\Bh(0)$ and $\Bh(\pi)$ are antisymmetric and non-singular but not necessarily real matrices. Since the manifold of antisymmetric non-singular matrices $\antisym$ is connected \cite{Zumino}, we can find a continuous connection from $\Bh(\pi)$ to $\Bh(0)$, which gives the antisymmetric closure $\Bh_c$ described in the proposition.

We have to show that $\wind(c)\ \mod2$ is independent of the antisymmetric closure we choose. So let $\Bh_c(k)$ and $\Bh_{c'}(k)$ be two different closures. By concatenating the two closure segments (see \Qref{fig:d3flyingcarpet}) we get a closed path in $\antisym$. The winding number of its determinant will be $\wind(c)-\wind(c')$. Hence we need to show that any closed path in $\antisym$ has even winding number. Indeed, for $A\in\antisym$ we have $\det A=\pfaff(A)^2$ where also $\pfaff(A)\neq0$. Hence, for a closed path $\wind\bigl(\det A(\cdot)\bigr)=2\wind\bigl(\pfaff A(\cdot)\bigr)\in2\Ir$.

For a continuous deformation of $\Bh$ the end points $\Bh(0)$ and $\Bh(\pi)$ in $\antisym$ move continuously, so the closures can also be made to change continuously by considering as deformed closures those obtained by joining the deformations of $\Bh(0)$ and $\Bh(\pi)$ to the original closure. Hence $\wind(c)$ is a homotopy invariant. It is also a complete invariant. To see this, we can first deform each walk to a flat band one. Moreover, we can choose the antisymmetric closure to be unitary, since for a flat-band walk the $\Bh(k)$ are unitary and the restriction of $\antisym$ to unitary elements remains connected \cite{Zumino}. We choose it so that $\wind(c)\in\{0,1\}$, and $\Bh_c(3\pi/2)=A$ for some fixed $A\in\antisym$. By including parts of the  antisymmetric closure as initial and final segments of $\Bh$ we deform to a path with $\Bh(0)=\Bh(\pi)=A$. In this normal form no closure operation is needed, and $\wind(c)$ is just the winding number of the determinant of a closed path in the unitary group. Since the fundamental group of the unitary group is $\Ir$, equality of $\wind(c)$ implies that two walks can be deformed into each other.

To get formula \eqref{DIIIwinding} it now suffices to check it on a generating example. The case of trivial index is straightforward: if $\sixR(W)=0$, $W(k)$ is homotopic to a trivial walk with $c(k)=c$ for all $k$ and therefore $2\ \wind(c)\equiv0\ \mod4$.\\
A non-trivial generating example for symmetry type $\symDIII{}$ is given by
	\begin{equation}
	\Bh(k)=\begin{pmatrix}
		0 & -e^{-i k}\\
		e^{ik} & 0
	\end{pmatrix},
	\end{equation}
	with $\ph'=K$ being the complex conjugation. A possible closure is then given by $-ie^{i k}\sigma_2$, where $\sigma_2$ denotes the second Pauli matrix. Then $\wind(c)=1$, in accordance with the claimed formula, since, by $W^2=-\idty$, $\sixR(W)=\ker\bigl(\im(PWP)\bigr)=\ker\bigl(-iPWP\bigr)=2$.
\end{proof}

For continuously differentiable bands, we then get an index formula with two contributions. A Berry type integral, stemming from the winding of the determinant of $\Bh(k)$ (compare \Qref{cor:chiralberry}) and the quotient of Pfaffians, resembling the invariant for \symD-walks (\Qref{prop:symDti}). For \symDIII-symmetric superconductors a formula similar to \eqref{DIIIberry} was derived in \cite{qi2010topological}.

\begin{cor}
	Let $W$ be a flat-band walk of symmetry type $\symDIII{}$. When $\Wh$ is continuously differentiable, $\sixR(W)$ can be written in a Berry phase type formula minus a correction term, i.\ e.
	\begin{equation}\label{DIIIberry}
	\sixR(W)\equiv\frac{2}{\pi i}\left(\int_0^\pi dk\sum_{\alpha=1}^d\Bigl\langle\phi_\alpha(k),\frac{d\phi_\alpha(k)}{dk}\Bigr\rangle-\log\left(\frac{\pf(\Bh(\pi))}{\pf(\Bh(0))}\right)\right)\quad\mod 4.
	\end{equation}
\end{cor}

\begin{proof}
	The winding integral of $c$ has two contributions, one from $\det(\Bh(k))$, for $k\in[0,\pi]$ and one from the antisymmetric closure for $k\in[\pi,2\pi]$. The first summand is the integral over the Berry connection, according to \Qref{cor:chiralberry}. We already showed, that, evaluated $\mod 2$, the second contribution is independent of which closure we choose. Hence it depends only on the endpoints and evaluates to the correction term in \eqref{DIIIberry}, if we use $\det(A)=\pf(A)^2$ and $\Bh(2\pi)=\Bh(0)$.
\end{proof}


\section{Decay properties of boundary eigenfunctions}\label{sec:decay}

In this section we consider a translation invariant walk $W$ with strictly finite propagation. Suppose we have joined it with another walk on the negative half axis, so we possibly get some eigenfunctions $\phi$ for eigenvalue $\pm1$.
We claim that in this case $\abs{\phi(x)}\leq\lambda^x$ as $x\to\infty$. Knowing the decay rate is crucial for applications of the theory to finite systems (see Sect. 9 in \cite{LongVersion}): When the bulk does not extend all the way to $+\infty$ but only has length $L$ we still expect to find eigenvalues near $\pm1$, whose distance from $\pm1$ is of the order $\lambda^L$.

Restricting, without loss, to the eigenvalues at $+1$, we have to characterize the solutions of the equation $(W-\idty)\phi(x)=0$ for $x\geq 0$, where we assumed the transition region to lie somewhere on the negative half-axis and does not intersect with $x=0$. This is a $\Cx^d$-valued linear recursion relation of finite order which, however, is not explicit usually: Using \eqref{Wti} we cannot simply solve for the highest occurring $\psi(y)$, because $W(x)$ need not be invertible for the lowest value of $x$ where this is non-vanishing. Nevertheless, one can usually find an appropriate selection of components of $\phi(x),\phi(x+1),\ldots,\phi(x+r)$ for which the recursion can be solved by iterating a fixed matrix. This is called the {\it transfer matrix} method. But does it always work? Consider, e.g.,  a walk in which one component of $\Cx^d$ is simply left invariant. Then this component in $\phi(x)$ drops out of the equation altogether, so we can say nothing about decay. In this case, we can also make eigenfunctions on the whole line, so the gap condition is violated. This means that we have to turn a spectral condition (existence of the gap) into an algebraic property (existence of a transfer matrix).

Even if we cannot directly turn the eigenvalue equation into a matrix iteration, we can try the exponential ansatz $\phi(x)=\lambda^x\phi_0$. Using \eqref{Wti}, the eigenvalue equation $W\phi(x)=\phi(x)$ then becomes
\begin{eqnarray} \label{wtfi}
   \Wt(\lambda)\phi_0&=&\phi_0 ,\qquad\text{where}\\
   \Wt(\lambda)&=& \sum_yW(y)\lambda^{-y}.
\end{eqnarray}
Comparison with \eqref{Whk} implies that $\Wh(k)=\Wt\bigl(e^{-ik}\bigr)$, so $\Wt$ is an analytic continuation of $\Wh$ from the unit circle to the unit disc. Clearly, a necessary condition for solving \eqref{wtfi} is $\det(\Wt(\lambda)-\idty)=0$, which is an algebraic equation for $\lambda$. The following proposition confirms that the strategy based on the exponential ansatz is indeed valid.

\begin{prop}\label{prop:expok}
Let $W$ be a strictly local translation invariant quantum walk with spectral gap at $1$. Let $\phi\in\HH$ satisfy $(W-\idty)\phi(x)=0$ for $x\geq0$. Then, there are vectors $\phi_{\lambda,i}\in\Cx^d$, and exponents $m_{\lambda,i}\in\Nl$ so that, for  $x\geq0$,
\begin{equation}\label{phixexpo}
  \phi(x)=\sum_\lambda\sum_i x^{m_{\lambda,i}}\lambda^x \ \phi_{\lambda,i},
\end{equation}
where $\lambda$ runs over the finite set of solutions of $\det\bigl(\Wt(\lambda)-\idty\bigr)=0$ with $\abs\lambda<1$, and, for each $\lambda$, $i$ runs from $1$ to the algebraic multiplicity of the zero.
\end{prop}

\begin{proof}
Let $\psi=(W-\idty)\phi$, which is a function vanishing for $x\geq0$. Let $\widetilde\psi(\lambda)=\sum_{x<0} \psi(x)\lambda^{-x}$, which is absolutely convergent and analytic for $\abs\lambda<1$. The boundary value $\widetilde\psi(e^{-ik})=\widehat\psi(k)$ is the Fourier transform of $\psi$.
Now pick some $x\geq0$ and a vector $\chi\in\Cx^d$, and let $\delta_x\otimes\chi\in\HH$ be the vector equal to $\chi$ at $x$ and zero otherwise. Then
\begin{eqnarray}
\brAket\chi{\phi(x)} &=& \brAket{\delta_x\otimes\chi}{(W-\idty)\inv\psi}
       =\int\frac{dk}{2\pi}e^{-ikx} \brAket{\chi}{(\Wh(k)-\idty)\inv\widehat\psi(k)} \\
       &=&\frac{i}{2\pi}\int\!du\ u^x
           \Bigl\langle{\chi}\Bigm|{(\Wt(u)-\idty)\inv\,\frac1u\widetilde\psi(u)}\Bigr\rangle,
\end{eqnarray}
where we have substituted $u=e^{-ik}$, and the integral is around the unit circle. Now $u^x$ and $u\inv\widetilde\psi(u)$ are analytic on the unit disc, and $(\Wt(u)-\idty)\inv$ is analytic except at the points where the operator $(\Wt(u)-\idty)$ becomes singular. These are the solutions of $\det\bigl(\Wt(u)-\idty\bigr)=0$, the order of the pole being at most the order of the algebraic multiplicity of this zero. Hence by the Residue Theorem, we can write the integral as a sum of evaluations of $u^x$, and possibly some derivatives of $u^x$, at $u=\lambda$.
\end{proof}

\subsubsection*{Example}

Let us confirm \Qref{prop:expok} with the split-step example (\Qref{sec:splitstep}). Eigenvectors of a decoupled version are determined by solving the eigenvalue equations $W\phi=s\phi$ and $\ch\phi=\chi\phi$ ($s,\chi=\pm1$) as a recursion relation in the bulk, giving exponential solutions, and then selecting those solutions satisfying the boundary conditions.
Let us sketch this explicitly for $s=1$, to keep the example concise:

Solving $\det\big(\Wt(\lambda)-\idty\big)=0$ (with $\Wt$ as in \Qref{prop:expok}) for the decay coefficients $\lambda$ results in exactly two solutions, one for each value $\chi$ of the chirality (both are non-degenerate), with $\theta_\pm = \frac 1 2 (\theta_1\pm\theta_2)$ \cite{AsboRak}:
\begin{equation}
\lambda_\chi = \frac{2 \cos\theta_-}{\cos\theta_- + \chi \sin\theta_+} - 1.
\end{equation}
Since $\lambda_+\lambda_-=1$, we know that there is always at most one candidate for an exponentially decaying solution to the right. \Qref{prop:expok} then yields the ansatz
\begin{equation}
\phi(x) = \lambda_\chi^x \phi_0, \quad x\geq 0,
\end{equation}
where $\phi_0=(1,\chi)$ is determined (up to normalization) by $\ch\phi=\chi\phi$. The splitting coin now decides, which of these potential solutions can be made to fit at the boundaries, e.g., by shifting this solution to the right and inserting a finite number of transition components to match the boundary conditions. In this example, the walk affects only its direct neighbours, hence there is at most one transition element.

Since we know that every eigenvector of a decoupled walk can be chosen to be localized on one side of the cut, we can choose $\phi(x)=0$ for all $x<0$, and check if $\phi$ fulfils the eigenvalue equation for each of the four differently decoupled walks. We note that in this simple example, inserting transition elements could never make a modified $\phi'$ solve an eigenvalue equation, that $\phi$ did not solve before.

This, together with the eigenvectors for the eigenvalue $s=-1$, analyzed on both sides of the cut, leads to \Qref{fig:harlequin}. Consistent with the general theory of our classification \cite{LongVersion}, each eigenvector found this way contributes its chirality $\chi$ to the corresponding indices $\six_\pm,\sixR,\sixL$, depending on whether they stem from $s=\pm1$ and whether they are exponentially decaying to the left or right (while vanishing on the other side). This is an explicit demonstration of bulk-boundary-correspondence.
A web-application allows to check these results dynamically \cite{sse}.

\section{Summary and Outlook}

We gave a homotopy classification of the systems with non-trivial index group considered in \cite{LongVersion} under the additional assumption of translation invariance, i.e. essentially gapped one-dimensional quantum walks with discrete symmetries which commute with lattice translations. In doing so we generalized the correspondence between strictly local walks and analytic band structures and proved that the continuity of the bands implies essentially locality.
This classification is complete within the class of translation invariant systems, i.e. two translation invariant walks obeying the same symmetries can be deformed into each other if and only if their invariants agree. For the symmetry types $\symS=\symD,\symBDI$ we found an additional invariant, which depends on the cell structure and trivializes after regrouping neighbouring cells or adding trivial systems. For each symmetry type with non-trivial index group, concrete index formulas were derived. In particular, the classification was obtained without referring to topological K-theory.

Possible directions for future research include the generalization of the results obtained here to higher dimensional lattices. For Hamiltonian systems, such a program leads to the now famous periodic table for the index groups \cite{kitaevPeriodic}. This corresponds to the classification of vector bundles over higher-dimensional Brillouin zones. Proving a bulk-boundary correspondence in higher dimensions, from analogy with the Hamiltonian case one expects to observe (directed) transport along edges if the edge is chosen along one of the principal directions of the lattice.

\section*{Acknowledgements}
We would all like to thank Alberto Gr\"unbaum for stimulating discussions.

C. Cedzich acknowledges partial support by the Excellence Initiative of the German Federal and State Governments (ZUK 81) and the DFG (project B01 of CRC 183).

T. Geib, C. Stahl and R. F. Werner acknowledge support from the ERC grant DQSIM, the DFG SFB 1227 DQmat, and the European project SIQS.

The work of L. Vel\'azquez is partially supported by the research projects MTM2014-53963-P and MTM2017-89941-P from the Ministry of Science and Innovation of Spain and the European Regional Development Fund (ERDF), and by Project E-64 of Diputaci\'on General de Arag\'on (Spain).

A. H. Werner thanks the Humboldt Foundation for its support with a Feodor Lynen Fellowship and the VILLUM FONDEN via the QMATH Centre of Excellence (Grant No. 10059).

The publication of this article was funded by the Open Access fund of the Leibniz Universit\"at Hannover. 

\bibliography{tphbib}

\end{document}